\documentclass[]{relaxed_system_lab}

\usepackage[utf8]{inputenc}
\usepackage[T1]{fontenc}
\usepackage{geometry}
\usepackage{amsmath,amssymb}  %
\usepackage{charter}

\usepackage[toc,page,header]{appendix}

\usepackage{algorithm}
\usepackage[noend]{algorithmic}
\usepackage{array}
\usepackage{wrapfig}

\usepackage{amsmath}
\usepackage[utf8]{inputenc} 
\usepackage[T1]{fontenc}    
\usepackage{hyperref}       
\usepackage{url}            
\usepackage{booktabs}       
\usepackage{amsfonts}       
\usepackage[most]{tcolorbox}
\usepackage{nicefrac}       
\usepackage{microtype}      
\usepackage{xcolor} 
\usepackage{amssymb}
\usepackage{caption}
\usepackage{xspace}
\usepackage{multirow}
\usepackage{xcolor}
\usepackage{graphicx} 

\usepackage{xspace}
\usepackage{multirow}
\usepackage{seqsplit}
\newcommand{\sys}{\textsc{Autopoiesis}\xspace}

\usepackage{tcolorbox}
\usepackage{mdframed}
\mdfdefinestyle{insightbox}{
  backgroundcolor=green!5!white,
  linecolor=green!40!black,
  linewidth=0.3pt,
  roundcorner=2pt,
  innerleftmargin=4pt,
  innerrightmargin=4pt,
  innertopmargin=3pt,
  innerbottommargin=3pt
}
\mdfdefinestyle{motivationbox}{
  backgroundcolor=blue!5!white,
  linecolor=blue!40!black,
  linewidth=0.3pt,
  roundcorner=2pt,
  innerleftmargin=4pt,
  innerrightmargin=4pt,
  innertopmargin=3pt,
  innerbottommargin=3pt
}
\mdfdefinestyle{codebox}{
  backgroundcolor=black!4!white,
  linecolor=black!40!black,
  linewidth=0.3pt,
  roundcorner=2pt,
  innerleftmargin=4pt,
  innerrightmargin=4pt,
  innertopmargin=3pt,
  innerbottommargin=3pt
}

\ifdefined\final
\usepackage[disable]{todonotes}
\else
\usepackage[textsize=tiny]{todonotes}
\fi

\usepackage{natbib}
\usepackage{latexsym}

\usepackage{url}
\usepackage{amssymb}
\usepackage[utf8]{inputenc}
\usepackage{microtype}
\usepackage{booktabs}
\usepackage{pifont} 
\usepackage{multirow}
\usepackage{makecell}
\usepackage{paralist}
\usepackage{xspace}
\usepackage{color}
\usepackage{xcolor}
\usepackage{colortbl}
\usepackage{adjustbox}
\usepackage{hyperref} 
\usepackage[edges]{forest}
\usepackage{tikz} 
\usepackage{caption}
\usepackage{amsfonts}

\hypersetup{
    colorlinks,
    linkcolor={blue!80!black},
    citecolor={blue!80!black},
}
\tikzset{
    root/.style =             {align=center, text width=1cm, rounded corners=3pt, line width=0.3mm, fill=gray!10, draw=gray!80, font=\small},
    demographic/.style =         {align=center, text width=1.8cm, rounded corners=3pt, line width=0.3mm, fill=blue!10, draw=blue!80, font=\footnotesize},
    demographic_work/.style =    {align=center, text width=10cm, rounded corners=3pt, line width=0.3mm, fill=blue!10, draw=blue!0, font=\footnotesize},
    character/.style =         {align=center, text width=1.8cm, rounded corners=3pt, line width=0.3mm, fill=red!10, draw=red!80, font=\footnotesize},
    character_work/.style =    {align=center, text width=10cm, rounded corners=3pt, line width=0.3mm, fill=red!10, draw=red!0, font=\footnotesize},
    personalization/.style =           {align=center, text width=1.8cm, rounded corners=3pt, line width=0.3mm, fill=cyan!10, draw=cyan!80, font=\footnotesize},
    personalization_work/.style =      {align=center, text width=10cm, rounded corners=3pt, line width=0.3mm, fill=cyan!10, draw=cyan!0, font=\footnotesize},
    risk/.style =         {align=center, text width=1.8cm, rounded corners=3pt, line width=0.3mm, fill=orange!10, draw=orange!80, font=\footnotesize},
    risk_work/.style =    {align=center, text width=10cm, rounded corners=3pt, line width=0.3mm, fill=orange!10, draw=orange!0, font=\footnotesize},
}

\usepackage{CJK}

\newtcolorbox{promptbox}[1][]{
  enhanced,
  breakable,
  colback=promptboxlightgray,
  colframe=promptboxblue!30,
  arc=8pt,
  boxrule=0.5pt,
  left=12pt,
  right=12pt,
  top=8pt,
  bottom=8pt,
  fonttitle=\bfseries,
  fontupper=\linespread{1.2}\selectfont,
  title=#1
}

\title{\sys: A Self-Evolving System Paradigm for LLM Serving Under Runtime Dynamics}

\author{Youhe Jiang$^1$$^*$, Ran Yan$^1$$^*$, You Peng$^1$, Wenshuang Li$^1$, Taiyi Wang$^2$, \\Fangcheng Fu$^3$$^\dagger$, Binhang Yuan$^1$$^\dagger$}

\affiliation{$^1$HKUST, $^2$Powersense Technology Limited, $^3$Shanghai Jiao Tong University}

\contribution{$^*$Equal contribution, $^\dagger$Corresponding author}

\abstract{
Modern Large Language Model (LLM) serving operates in highly volatile environments characterized by severe runtime dynamics, such as workload fluctuations and elastic cluster autoscaling. Traditional serving systems rely on static, human-engineered serving policies (e.g., scheduling algorithms and rescheduling strategies) to manage these dynamics. However, these policies must navigate deeply intertwined runtime trade-offs (e.g., scheduling overhead vs. execution efficiency, rescheduling frequency vs. reconfiguration overhead), whose optimal balance is workload-specific and shifts continuously as runtime conditions evolve, rendering any fixed policy fundamentally unable to adapt.
We propose \sys, a novel \textit{online} \textit{self-evolving} system that shifts LLM serving from static policy deployment to continuous online policy evolution. \underline{First}, \sys introduces an \textit{LLM-driven program synthesis workflow} to evolve serving policies with respect to real-time observed dynamics, where the evolved policies reflect the optimal decision in navigating the complex, multi-dimensional trade-off space. 
\underline{Second}, \sys enables this synthesis process to operate continuously during serving, observing real-world system behavior, and rewriting the policy code as runtime trade-offs shift, thereby transforming policy design from a one-time offline endeavor into an ongoing system component, enabling autonomous adaptation to evolving runtime conditions. 
Together, we establish a new paradigm: Serving policies are no longer static artifacts designed by humans before deployment, but living code that LLMs continuously evolve throughout deployment to navigate runtime trade-offs beyond human design. We evaluate \sys across diverse runtime dynamics and show up to 53\% and on average 34\% improvements over state-of-the-art LLM serving systems.
}

\begin{document}

\maketitle

\section{Introduction}
\label{sec:intro}

The rapid proliferation of Large Language Model (LLM) serving has exposed an essential operational challenge: \textit{production serving paradigm is inherently dynamic.} User request patterns fluctuate unpredictably across time, with bursty arrivals, varying prompt lengths, and diverse output generation demands~\cite{wang2025burstgpt,sun2024llumnix,agrawal2024taming,li2023alpaserve,sheng2024fairness,zhong2024distserve}. Cluster-level resources change as autoscaling mechanisms add or remove compute nodes in response to load~\cite{fu2024serverlessllm,xiang2025aegaeon,wu2024loongserve,wagenlander2024tenplex}, while spot instance fluctuations alter cluster membership and hardware failures intermittently take nodes offline~\cite{miao2024spotserve,jiang2025thunderserve,mao2025skyserve}. These dynamics are ubiquitous in modern LLM serving~\cite{kwon2023efficient,yu2022orca}, where any serving system must handle via continuous deployment.

To accommodate such dynamics, serving systems employ \textit{serving policies} (e.g., scheduling algorithms and rescheduling strategies) that produce different \textit{serving plans} (including workload assignment, resource allocation, and parallelism strategy), in response to dynamic status. However, implementing an effective serving policy requires navigating deeply intertwined \textbf{trade-offs}---when runtime status shifts, the serving policy must decide whether to reschedule, how thoroughly to replan, and how aggressively to reconfigure: (\textbf{\underline{i}}) Rescheduling more frequently keeps plans fresh but accumulates overhead that may outweigh the benefit~\cite{sun2024llumnix,xiang2025aegaeon,agrawal2024taming}. (\textbf{\underline{ii}}) Investing more computation in scheduling yields better plans but extends the window of serving under a stale plan~\cite{sun2024llumnix,jiang2025thunderserve,qiu2024efficient}. (\textbf{\underline{iii}}) Reconfiguring to a new plan improves throughput but incurs transient degradation proportional to the magnitude of the change~\cite{miao2024spotserve,jiang2025thunderserve,cheng2024kunserve}. These trade-offs are tightly coupled, which results in a multi-dimensional trade-off space whose optimal operating point is workload-specific and shifts continuously as runtime conditions evolve.

\begin{figure}
    \centering
    \includegraphics[width=0.5\linewidth]{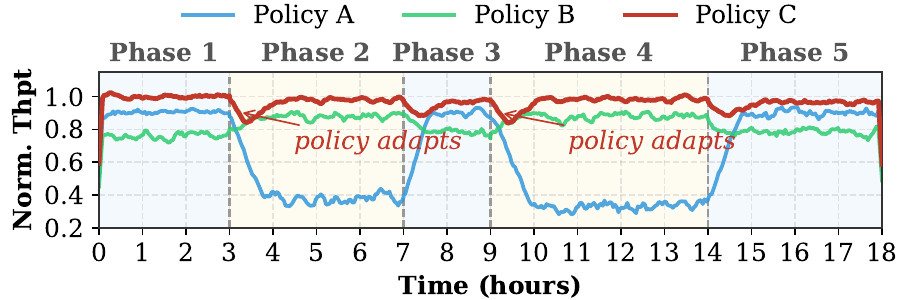}
    \caption{\small{Impact of workload dynamics on serving policy performance across five phases: Steady (Phases 1, 3, 5) and bursty (Phases 2, 4). Policy A uses thorough scheduling with aggressive reconfiguration, excelling under stability but collapsing under bursts. Policy B uses reactive scheduling with lightweight reconfiguration, thriving during bursts but underperforming in steady state. Policy C continuously adapts to shifting conditions, navigating optimal trade-offs and maintaining high throughput across all phases.}}
    \label{fig:runtime}
\end{figure}

Existing LLM serving systems~\cite{zhong2024distserve,wu2024loongserve,qin2024mooncake,sun2024llumnix} employ static, human-engineered serving policies\footnote{For example, DistServe~\cite{zhong2024distserve} replicates a pre-computed optimal parallelism configuration to scale throughput, exploiting the fact that all GPUs share identical capacity in a homogeneous cluster. HexGen~\cite{jiang2025hexgen} constrains each tensor-parallel group to use GPUs within a single node, since cross-node communication bandwidth is typically limited in heterogeneous clusters.}. While these systems generate different serving plans in response to runtime dynamics, the policies that produce those plans remain fixed, failing to account for the runtime trade-offs outlined above. As we demonstrate in~\autoref{fig:runtime}, this leads to significant performance degradation. We argue that this gap is inherent to the current system paradigm, due to two fundamental \textbf{limitations}: (\textbf{\underline{i}}) Human-engineered policies struggle to discover trade-off balances precisely tailored to specific runtime conditions, as the multi-dimensional trade-off space is large and deeply intertwined for human intuition to navigate. (\textbf{\underline{ii}}) The policies remain fixed throughout deployment, so even a well-tuned policy becomes suboptimal as conditions shift. Offline precomputation of policies for all possible runtime scenarios is not sufficient either, as the space of possible workload patterns, cluster configurations, and their temporal interactions is impossible to enumerate in advance. 
Static, human-engineered policies cannot navigate a trade-off space that is both complex and dynamic to solve once.

To navigate runtime dynamics, we identify two key requirements for an effective LLM serving paradigm:

\vspace{0.25em}
\noindent \textbf{\underline{Requirement 1}}: \textit{Automated trade-off exploration.} The system must be capable of exploring the multi-dimensional trade-off space and discovering policies precisely tailored to the current workload and cluster conditions, rather than relying on human intuition to manually balance the intertwined trade-offs of scheduling thoroughness, reconfiguration aggressiveness, and rescheduling frequency. This directly addresses limitation (\textbf{\underline{i}}), as the required trade-off balances are unattainable through human engineering.

\vspace{0.25em}
\noindent \textbf{\underline{Requirement 2}}: \textit{Continuous online evolution.} The system must continuously observe real-world system behavior and rewrite the policy code as runtime trade-offs shift. Rather than treating policy optimization as a one-time offline effort, the system should integrate the policy exploration process as an always-on component of the serving system, tracking shifts in workload and cluster conditions and adapting the policy accordingly throughout deployment. This directly addresses limitation (\textbf{\underline{ii}}), as the policy no longer remains static but evolves alongside the runtime conditions it governs.

LLM-driven program synthesis offers a compelling opportunity to fulfill both requirements simultaneously. Recent advances have demonstrated that combining the code-generation capabilities of LLMs with the iterative refinement of evolutionary algorithms enables automatic discovery of optimal solutions in design spaces far more vast than human engineers can navigate~\cite{romera2024mathematical,novikov2025alphaevolve}. This mechanism has successfully synthesized novel, highly-tailored algorithms across diverse domains, from mathematical and scientific optimization~\cite{liu2026evox,sharma2025openevolve} to hardware-specific GPU kernel tuning~\cite{liao2025kernelevolve,cheng2025barbarians}. In this work, we bring this synthesis capability to LLM serving, utilizing it to continuously explore and navigate the multi-dimensional trade-off space of runtime dynamics.

To explore this, we propose \sys, a \textit{novel}, \textit{self-evolving} paradigm that shifts LLM serving from static, human-engineered policy design to continuous, LLM-driven policy evolution. Our contributions are summarized as:

\vspace{0.25em}
\noindent \textbf{\underline{Contribution 1:}} We propose an \textit{LLM-driven program synthesis workflow} for serving policy optimization. The workflow combines an evolutionary algorithm that iteratively generates, mutates, and selects policy candidates; a highly calibrated, lightweight evaluator that faithfully reproduces real-world runtime dynamics for efficient candidate assessment; and a synthesis configuration that orchestrates the evolutionary synthesis process, including population management, mutation strategy, and feedback mechanism.

\vspace{0.25em}
\noindent \textbf{\underline{Contribution 2:}} We design and implement \sys, a self-evolving serving paradigm built upon a decoupled \textit{two-plane architecture} (i.e., data- and control-plane). The \textit{data plane} executes the current serving policy to produce serving plans for live deployment, while the asynchronous \textit{control plane} continuously leverages the LLM-driven program synthesis workflow to optimize the serving policy itself, rewriting the code as runtime conditions evolve. 

\vspace{0.25em}
\noindent \textbf{\underline{Contribution 3:}} We conduct a comprehensive evaluation of \sys under diverse runtime dynamics, including workload fluctuations and spot instance variations. Experimental results demonstrate that \sys achieves up to 53\% and on average 34\% improvement in system performance over state-of-the-art LLM serving systems.

\vspace{0.25em}
\noindent \textbf{\underline{Contribution 4:}} We provide a detailed system analysis showing how evolved policies navigate runtime trade-offs across diverse serving scenarios and how optimal trade-off balances shift with workload and cluster conditions, offering insights into the effectiveness and generalizability of our synthesis workflow for system optimization problems.

\section{Background and Related Work}
\label{sec:background}

\subsection{LLM Serving Preliminaries}
\label{subsec:llm serving preliminaries}
Serving production-scale LLM requests requires allocating substantial memory and computationally intensive workloads across large scale GPU clusters. Efficient management of this execution necessitates the continuous orchestration of several deeply intertwined components:

\vspace{0.25em}
\noindent \textbf{Serving plans.} A serving plan represents the concrete execution state of the cluster at any given moment. Formally, it is composed of three primary dimensions: (\textbf{\underline{i}}) \textit{Workload assignment}, which determines the mapping of heterogeneous user requests to specific models; systems such as Llumnix~\cite{sun2024llumnix} and DistServe~\cite{zhong2024distserve} optimize request assignment to mitigate load imbalance and enhance overall serving throughput. (\textbf{\underline{ii}}) \textit{Resource allocation}, which determines the provisioning of physical hardware to models; frameworks such as AlpaServe~\cite{li2023alpaserve}, Mooncake~\cite{qin2024mooncake}, and Aegaeon~\cite{xiang2025aegaeon} optimize hardware provisioning across concurrently served models to maximize cluster utilization and system capacity. (\textbf{\underline{iii}}) \textit{Parallelism strategy}, which determines the partitioning of model weights and computations across allocated hardware through techniques such as tensor~\cite{shoeybi2019megatron}, pipeline~\cite{huang2019gpipe}, data~\cite{li2023alpaserve}, sequence~\cite{jacobs2023deepspeed}, and expert parallelism~\cite{rajbhandari2022deepspeed}; systems such as LoongServe~\cite{wu2024loongserve} and Tenplex~\cite{wagenlander2024tenplex} optimize parallelism configurations to maximize hardware efficiency. Crucially, these dimensions cannot be determined in isolation; they must be tightly co-optimized. 

\vspace{0.25em}
\noindent \textbf{Serving policies.} While a serving plan represents a static snapshot of cluster execution, a serving policy constitutes the active algorithmic logic that continually evaluates and updates this plan. A serving policy comprises two primary components: (\textbf{\underline{i}}) \textit{Scheduling algorithm}, which serves as the core optimization engine of the system; it takes workload, model, and cluster information as input and generates an optimal serving plan that satisfies a specified system objective (e.g., SLO attainment). Frameworks such as AlpaServe~\cite{li2023alpaserve}, DistServe~\cite{zhong2024distserve}, Aegaeon~\cite{xiang2025aegaeon}, and SkyServe~\cite{mao2025skyserve} deploy sophisticated scheduling algorithms to determine the optimal workload assignment and model deployment across complex hardware topologies. (\textbf{\underline{ii}}) \textit{Rescheduling strategy}, which serves as the temporal control mechanism that governs the dynamic nature of the system; it determines when (i.e., the trigger condition) and how (i.e., the target plan) the system should reconfigure to adapt to shifting workload or cluster conditions. Frameworks such as Llumnix~\cite{sun2024llumnix}, ThunderServe~\cite{jiang2025thunderserve}, and SpotServe~\cite{miao2024spotserve} employ adaptive rescheduling logic to dynamically migrate request and model states and rapidly adjust cluster boundaries in response to sudden load imbalances and unexpected hardware preemptions.

\vspace{0.25em}
\noindent \textbf{Limitations.} Despite the sophistication of existing frameworks, current approaches for policy design exhibit three key limitations: \underline{First}, they rely on human-engineered heuristics and case-specific constraints; a policy tuned for a particular workload distribution or hardware topology rarely generalizes to others. \underline{Second}, these manually designed policies lack a holistic mechanism to navigate the deeply intertwined runtime trade-offs detailed in~\S\ref{sec:intro}, such as balancing scheduling thoroughness against execution efficiency, or weighing rescheduling frequency against the latency cost of cluster reconfiguration. \underline{Third}, these policies remain entirely static throughout the serving phase. Because runtime dynamics continuously shift the optimal balance of these trade-offs, a fixed policy inevitably becomes suboptimal as conditions evolve, leaving the system unable to adapt its own governing logic to maintain performance under real-world volatility.

\subsection{LLM-Driven Program Synthesis}
\label{subsec:llm-driven program synthesis}
LLM-driven program synthesis is an emerging paradigm that combines the code-generation capabilities of LLMs with the iterative refinement of evolutionary algorithms to automatically discover high-performing programs.

\vspace{0.25em}
\noindent \textbf{LLM-driven program synthesis.} The LLM-driven program synthesis includes four primary stages:
(\textbf{\underline{i}}) \textit{Seed initialization and program representation} determines the starting population and the level of abstraction at which evolution operates; e.g., AlphaEvolve~\cite{novikov2025alphaevolve} and CodeEvolve~\cite{assumpccao2025codeevolve} support codebase-scale evolution spanning multiple functions and programming languages.
(\textbf{\underline{ii}}) \textit{LLM-driven variation} replaces the hand-crafted mutation and crossover operators of classical genetic programming~\cite{koza1994genetic} with the semantic code-transformation capabilities of LLMs; AlphaEvolve~\cite{novikov2025alphaevolve} employs an LLM ensemble with structured diff-based modifications for precise, localized code changes, while EoK~\cite{chen2025evolution} guides LLM-driven mutation with RAG-enriched prompts that incorporate actionable optimization thoughts mined from established kernel libraries' development histories.
(\textbf{\underline{iii}}) \textit{Automated evaluation} assesses each candidate against a well-defined fitness criterion, serving as the critical safeguard against LLM hallucinations; KernelEvolve~\cite{liao2025kernelevolve} integrates correctness validation, performance benchmarking, and hardware profiling into a unified evaluation pipeline across heterogeneous accelerators.
(\textbf{\underline{iv}}) \textit{Selection and population management} governs the exploration--exploitation balance by determining which candidates survive to inform future generations; OpenEvolve~\cite{sharma2025openevolve} adopts MAP-Elites-inspired~\cite{mouret2015illuminating} program databases that preserve diverse high-quality candidates across a discretized feature space.

\vspace{0.25em}
\noindent \textbf{Accommodating dynamics in LLM serving.} While existing LLM-driven evolutionary frameworks have achieved remarkable success, applying them to the runtime dynamics of LLM serving requires overcoming two key challenges.
\underline{First}, \textit{capturing runtime trade-offs}: Candidate evaluation in standard evolutionary processes fails to capture how policy decisions physically impact system execution. Addressing this requires a domain-specific evolutionary workflow: An evaluation methodology that faithfully reproduces real-world conditions to measure deeply intertwined runtime trade-offs (e.g., reconfiguration overhead vs. throughput gains), coupled with a synthesis configuration that actively steers LLM mutations to navigate this multi-dimensional space.
\underline{Second}, \textit{evolving with runtime dynamics}: Offline program synthesis produces static policies optimized for a snapshot of conditions, which are unable to adapt as those conditions shift. Addressing this requires rearchitecting the synthesis paradigm into a continuous, online control loop that monitors live cluster state, grounds fitness evaluation in real-time telemetry, and deploys evolved policy code in live serving.

\section{Motivation}
\label{sec:motivation}

We conduct two empirical studies on real-world subsampled workload traces~\cite{yao2026opentela} to motivate the design of \sys (trace information detailed in~\autoref{appendix:trace}). The first motivation demonstrates that finding the optimal runtime trade-off balance is critical to serving performance. The second motivation demonstrates that this optimal balance shifts as runtime conditions evolve, necessitating continuous re-exploration. Each study is structured as a characterization of the phenomenon, an analysis of its root cause, and a motivation for the design requirement of \sys.

\begin{figure}
    \centering
    \includegraphics[width=0.5\linewidth]{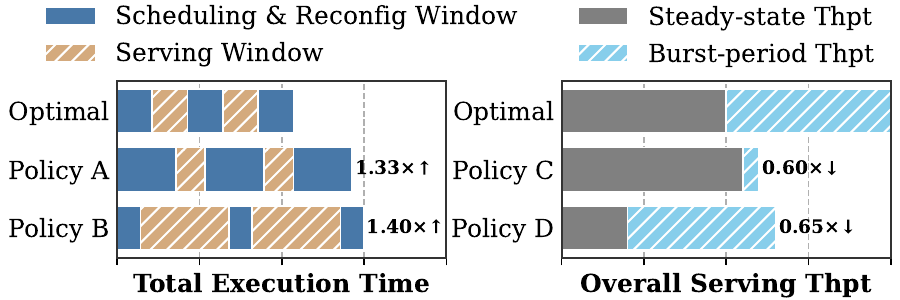}
    \caption{\small{\textbf{\underline{Left:}} Runtime trade-off navigation: Policies~A (maximum scheduling and reconfiguration intensity) and B (minimum intensity) compared to an optimal oracle. \textbf{\underline{Right:}} Shifting trade-off adaptation: Policies~C (optimized for steady-state) and D (optimized for burst periods) compared to a dynamically adaptive baseline.}}
    \label{fig:motivation}
\end{figure}

\vspace{0.25em}
\noindent \textbf{\underline{Characterization 1}}: \textit{Navigating runtime trade-offs is critical to serving performance.}
We evaluate LLM serving across a workload trace spanning three rescheduling intervals, examining how scheduling thoroughness (i.e., the computational budget allocated to generating serving plans) and reconfiguration aggressiveness (i.e., the volume of system state reconfigured to achieve optimal placement) jointly affect total execution time. As shown in~\autoref{fig:motivation} (left), we compare two extreme policies against an optimal oracle. Policy~A maximizes both scheduling thoroughness and reconfiguration aggressiveness. While this minimizes the actual serving time, the extended scheduling and reconfiguration window forces the cluster to operate under a stale plan in the interim, increasing total execution time by 33\%. Conversely, Policy~B minimizes scheduling and reconfiguration overhead via a lightweight greedy algorithm. However, the resulting suboptimal plans lead to inefficient resource utilization, increasing total execution time by 40\%. Neither extreme achieves competitive performance; the optimal operating point lies in a narrow, workload-specific region between them.

\vspace{0.25em}
\noindent \textbf{Analysis.} \textit{The runtime trade-off space is too complex for human engineering to navigate.}
While the above characterization establishes that finding the right trade-off balance is critical, identifying this balance through human engineering faces compounding difficulty. First, the relationship among scheduling thoroughness, reconfiguration aggressiveness, and serving throughput is highly nonlinear, requiring exhaustive trial and error to locate even a single effective configuration. Second, the optimal balance is strictly deployment-specific: Any variation in hardware topology or workload distribution shifts the trade-off surface, invalidating prior design choices and demanding complete recalibration. Third, while the above analysis isolates
only two variables, production environments require jointly co-optimizing these alongside rescheduling frequency, causing the design space to expand exponentially. These compounding challenges make manual policy engineering struggle to discover optimal trade-off balances across diverse deployment conditions.

\vspace{0.5em}
\begin{mdframed}[style=motivationbox]
\small
\noindent \textbf{Motivation 1: LLM-driven policy synthesis (\S\ref{sec:intro}, Requirement~1).} The complexity and deployment-specificity of the runtime trade-off space demand an approach that can systematically explore the multi-dimensional space and discover policies precisely tailored to the current conditions---capabilities that exceed the reach of manual engineering. This motivates the design of an LLM-driven program synthesis workflow that evolves serving policies with respect to real-world runtime dynamics.
\end{mdframed}
\vspace{0.5em}

\vspace{0.25em}
\noindent \textbf{\underline{Characterization 2}}: \textit{Continuous adaptation to shifting runtime trade-offs is necessary.}
We further evaluate LLM serving under a dynamic workload trace that alternates between steady-state request arrivals (i.e., low workload variation) and sudden traffic bursts (i.e., high workload variation). As shown in~\autoref{fig:motivation} (right), we compare two statically tuned policies against an optimally adaptive baseline. Policy~C, tuned for the optimal trade-off during steady-state operations, achieves peak throughput during stable phases but collapses under traffic bursts, degrading end-to-end throughput by 40\% due to severe queuing delays. Conversely, Policy~D, tuned for the optimal trade-off during burst periods, seamlessly absorbs spikes but yields suboptimal plans during steady periods, leaving hardware underutilized and reducing throughput by 35\%. No single static policy sustains optimal performance when runtime conditions shift.

\vspace{0.25em}
\noindent \textbf{Analysis}. \textit{Static policy optimization is insufficient under runtime dynamics.}
Even if Motivation~1 is addressed, where the system can discover an optimal policy for the given conditions, the above characterization shows that such a policy becomes suboptimal the moment runtime trade-offs shift. The root cause is that static, offline policy design presupposes a stationary operating environment with a fixed optimal balance point. Once conditions change (e.g., transitioning from steady-state to burst traffic), the hardcoded trade-off balance hardly matches the new optimum. Crucially, such fluctuations are not isolated anomalies but the continuous, standard operating reality of production LLM serving, rendering any one-time policy optimization insufficient.

\vspace{0.5em}
\begin{mdframed}[style=motivationbox]
\small
\noindent \textbf{Motivation 2: Continuous online evolution (\S\ref{sec:intro}, Requirement~2).} Since the optimal trade-off balance shifts continuously with runtime dynamics, the policy synthesis process itself must operate continuously during serving rather than as a one-time offline effort. This motivates the design of a self-evolving serving paradigm that integrates LLM-driven policy synthesis as an always-on component, observing real-world system behavior and rewriting the policy code as runtime conditions evolve.
\end{mdframed}
\vspace{0.5em}

\section{\sys Overview}
\label{sec:overview}

To navigate the deeply intertwined and continuously shifting runtime trade-offs of modern serving environments, \sys employs a two-plane system architecture that decouples \textit{policy execution} from \textit{policy evolution}. By separating these two concerns, the architecture ensures that runtime trade-offs are thoroughly explored and that trade-off shifts are properly handled without disrupting live serving.

\vspace{0.25em}
\noindent \textbf{Two-plane system architecture.} The architecture comprises two decoupled planes for a closed loop of adaptation:
\begin{itemize}
\vspace{-0.25em}
    \item \textbf{The data plane (policy execution):} This plane handles live request traffic by executing the active serving policy. Based on real-time workload and cluster conditions, it continuously generates and updates serving plans, including workload assignment, resource allocation, and parallelism strategies. To capture the runtime trade-off shifts discussed in~\S\ref{sec:motivation}, the data plane periodically takes a snapshot of the recent runtime trace (i.e., a sliding window of request arrival patterns and cluster resource availability) and transmits it to the control plane.
    \item \textbf{The control plane (policy evolution):} This plane operates asynchronously to optimize the serving policy without interrupting live serving. Upon receiving each snapshot, it launches the LLM-driven program synthesis workflow to evolve the serving policy (i.e., rewrite the policy code). The workflow explores the multi-dimensional trade-off space (e.g., scheduling thoroughness vs.\ serving performance) relative to the snapshotted runtime trace to derive an optimized policy. Once a superior policy is found, the control plane deploys it to the data plane.
    \vspace{-0.25em}
\end{itemize}

\begin{figure}
    \centering
    \includegraphics[width=0.5\linewidth]{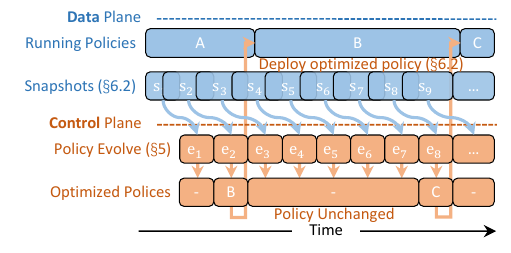}
    \caption{\small{Overview of \sys's two-plane system architecture. The data plane executes the serving policy and transmits snapshotted runtime trace to the control plane, while the control plane continuously evolves the policy and deploys it to the data plane when a superior policy is found, forming a self-evolving loop.}}
    \label{fig:workflow}
\end{figure}

\vspace{0.25em}
\noindent \textbf{The self-evolving loop.} As shown in \autoref{fig:workflow}, these two planes transform LLM serving into a self-evolving process. 
This architecture ensures that the system does not merely react to dynamics by altering its \textit{plan}, but proactively adapts its \textit{policy} to maintain optimal serving performance as the runtime trade-offs shift. We detail the LLM-driven program synthesis workflow that powers this exploration in~\S\ref{sec:llm-driven} and the implementation of \sys in~\S\ref{sec:impl}.

\section{LLM-Driven Program Synthesis}
\label{sec:llm-driven}

This section presents the LLM-driven program synthesis workflow that powers \sys's control plane. We begin by formalizing the runtime trade-off navigation problem that the workflow must solve (\S\ref{subsec:formulation}), then provide an overview of the synthesis workflow (\S\ref{subsec:workflow}), followed by a detailed description of the evaluator (\S\ref{subsec:evaluator}) and the synthesis configuration (\S\ref{subsec:synthesis-config}).

\subsection{Problem Formulation}
\label{subsec:formulation}

\vspace{0.25em}
\noindent \textbf{Objective.}
Given a runtime trace that captures evolving workload and cluster conditions (i.e., the snapshotted runtime trace mentioned in~\S\ref{sec:overview}), \sys seeks to evolve a serving policy that minimizes the end-to-end trace completion time $\text{T}_{\text{total}}$.

\vspace{0.25em}
\noindent \textbf{Continuous execution constraint.}
As is standard in online LLM serving~\cite{sun2024llumnix,miao2024spotserve,jiang2025thunderserve}, the system operates continuously: It does not pause during scheduling or reconfiguration. When a rescheduling event is triggered, the scheduler computes a new serving plan while the cluster continues serving under the previous (now stale) plan. Once the new plan is ready, the cluster reconfigures to it; during reconfiguration, the overlapping portion of the serving plan continues serving, while the non-overlapping portion is being transitioned, resulting in degraded throughput proportional to the reconfiguration magnitude. This constraint is the cause of the three runtime trade-offs identified in~\S\ref{sec:intro}: As serving never pauses, every scheduling decision and every reconfiguration action has a direct, concurrent impact on serving performance, coupling these costs into the trade-off space that must be navigated.

\vspace{0.25em}
\noindent \textbf{Serving policy formulation.}
Under the continuous execution constraint, navigating the runtime trade-off space reduces to two decisions: \textit{When} to reschedule and \textit{how} to reschedule. We formalize these as a pair of co-evolved functions that together constitute the serving policy:
\begin{itemize}
    \item $\texttt{should\_reschedule}(\texttt{ctx}) \rightarrow \{0,1\}$, the rescheduling strategy that determines whether current conditions have shifted sufficiently to justify rescheduling. This controls the \seqsplit{rescheduling} frequency $\text{N}$ within the given trace (Trade-off (\textbf{\underline{i}})).
    \item $\texttt{schedule}(\texttt{ctx}) \rightarrow \texttt{plan}$, the scheduling algorithm that produces a new serving plan, which controls both the scheduling overhead (Trade-off (\textbf{\underline{ii}})) and the reconfiguration magnitude of each new plan relative to the old plan. (Trade-off (\textbf{\underline{iii}})).
\end{itemize}
Both functions receive a shared context $\texttt{ctx}$, a state observation captured at each monitoring point during the trace. Each observation contains the workload and cluster conditions at that moment, along with the currently deployed plan (or $\varnothing$ at cold start). For instance, when request arrival rates surge at a particular point in the trace, $\texttt{should\_reschedule}$ evaluates through its internal logic whether the shift is significant enough to incur a rescheduling event; if rescheduling is triggered, $\texttt{schedule}$ determines the new serving plan. These two functions jointly dictate every scheduling and reconfiguration decision throughout the trace, and their behavior directly determines the end-to-end trace completion time $\text{T}_{\text{total}}$ as defined by the following execution model.

\begin{figure}
    \centering
    \includegraphics[width=0.5\linewidth]{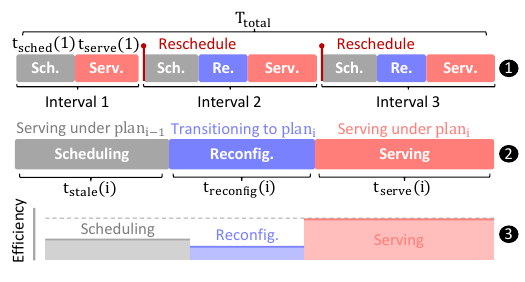}
    \caption{\small{Execution model illustration. \textbf{\textcircled{1}} Rescheduling intervals composing $\text{T}_{\text{total}}$: The first interval is a cold start, while subsequent intervals are triggered by $\texttt{should\_reschedule}$ and each contains three phases. \textbf{\textcircled{2}} Detailed view of an interval ($\text{i} \geq 2$) with three phases: Scheduling, reconfiguration, and serving. \textbf{\textcircled{3}} Example runtime efficiency profile: Efficiency is degraded during scheduling and reconfiguration, and maximized during serving.}}
    \label{fig:execmodel}
\end{figure}

\vspace{0.25em}
\noindent \textbf{Execution model.}
We now formalize the cost structure that arises when these two functions execute under the continuous execution constraint. We illustrate the intervals and phases of the execution model in~\autoref{fig:execmodel}. The first rescheduling interval is a cold start: $\texttt{schedule}$ computes $\texttt{plan}_1$, incurring scheduling cost $\text{t}_{\text{sched}}(1)$ (including the model initialization cost), and the system then serves under it, incurring serving cost $\text{t}_{\text{serve}}(1)$. Each subsequent interval $\text{i} \geq 2$, triggered when $\texttt{should\_reschedule}$ returns $\texttt{true}$ at a monitoring point, follows a three-phase pattern:
\begin{itemize}
    \item \textbf{Phase 1: Scheduling.} $\texttt{schedule}(\texttt{ctx}_\text{i})$ generates the new plan $\texttt{plan}_\text{i}$; meanwhile, the system continues serving under previous plan $\texttt{plan}_{\text{i}-1}$ at degraded efficiency due to the workload-plan mismatch. The stale serving cost $\text{t}_{\text{stale}}(\text{i})$ corresponds to this scheduling period on the timeline.
    \item \textbf{Phase 2: Reconfiguration.} The system transitions from $\texttt{plan}_{\text{i}-1}$ to $\texttt{plan}_\text{i}$; meanwhile, the overlapping portion of the serving plan continues serving at reduced efficiency as the non-overlapping portion is being reconfigured (e.g., migrating KV caches, redistributing model weights, or adjusting parallelism configurations). The reconfiguration serving cost $\text{t}_{\text{reconfig}}(\text{i})$ corresponds to this reconfiguration period on the timeline.
    \item \textbf{Phase 3: Serving.} The system serves under the new plan $\texttt{plan}_\text{i}$ at full efficiency until the next rescheduling trigger; all resources are now aligned with the current workload. The serving cost $\text{t}_{\text{serve}}(\text{i})$ corresponds to this stable serving period on the timeline.
\end{itemize}
The total trace completion time is therefore:
\vspace{-0.25em}
\begin{equation}
\footnotesize
\label{eq:ttotal}
\text{T}_{\text{total}} = \text{t}_{\text{sched}}(1) + \text{t}_{\text{serve}}(1) + \sum_{\text{i}=2}^{\text{N}} \Big[ \text{t}_{\text{stale}}(\text{i}) + \text{t}_{\text{reconfig}}(\text{i}) + \text{t}_{\text{serve}}(\text{i}) \Big]%
\end{equation}%

\vspace{-0.25em}
\noindent where $\text{N}$ is not a fixed constant but determined by how often $\texttt{should\_reschedule}$ triggers rescheduling during the trace, and each cost component within the summation reflects the behavior of $\texttt{schedule}$ at the corresponding interval. The objective of our program synthesis workflow is to evolve the algorithmic implementation of both $\texttt{should\_reschedule}$ and $\texttt{schedule}$, including their decision thresholds and computational strategies, to jointly minimize $\text{T}_{\text{total}}$. As these two functions govern every runtime decision during serving, optimizing their implementation directly determines the system's runtime behavior. Notably, this execution model naturally generalizes to diverse runtime dynamics, including workload fluctuations, cluster autoscaling, and spot instance preemptions, as any such event is reflected in $\texttt{ctx}$ and processed through the same trigger-schedule-reconfigure pipeline. Moreover, the cost components within each phase can be calibrated to match any specific system implementation (e.g., the specialized reconfiguration mechanism in SpotServe~\cite{miao2024spotserve}), making the formulation portable across different serving frameworks.

\vspace{0.25em}
\noindent \textbf{Trade-off space.} However, minimizing $\text{T}_{\text{total}}$ is not straightforward: The cost components in Equation~\ref{eq:ttotal} (i.e., $\text{t}_{\text{stale}}\allowbreak, \text{t}_{\text{reconfig}}\allowbreak, \text{t}_{\text{serving}}$) are interdependent, and improvements along one axis frequently come at the expense of another, creating a multi-dimensional trade-off space that \texttt{should\_reschedule} and \texttt{schedule} must jointly navigate. The trade-offs characterized empirically in~\S\ref{sec:motivation} are examples of this broader space:
\begin{itemize}
\item \textit{Trade-off (\textbf{\underline{i}}): Rescheduling frequency vs.\ per-interval overhead.} Frequent rescheduling (large $\text{N}$) keeps plans fresh but accumulates scheduling and reconfiguration overhead; infrequent rescheduling (small $\text{N}$) avoids this overhead but permits serving under suboptimal plans.

\item \textit{Trade-off (\textbf{\underline{ii}}): Scheduling thoroughness vs.\ stale serving cost.} Thorough scheduling produces higher-quality plans (lower $\text{t}_{\text{serve}}$) but extends the stale-plan serving window (higher $\text{t}_{\text{stale}}$); lightweight scheduling minimizes stale cost but may degrade plan quality.

\item \textit{Trade-off (\textbf{\underline{iii}}): Reconfiguration aggressiveness vs.\ reconfiguration overhead.} Aggressive reconfiguration achieves a globally optimal plan (lower $\text{t}_{\text{serve}}$) but incurs prolonged transition degradation (higher $\text{t}_{\text{reconfig}}$); conservative reconfiguration minimizes transition cost but yields limited serving improvement.
\end{itemize}
To provide concrete intuition for how these trade-offs manifest in practice, we present a detailed analysis in~\S\ref{sec:analysis}, illustrating how the optimal runtime decisions differ across diverse serving scenarios such as workload fluctuations~\cite{sun2024llumnix}, spot instance preemptions~\cite{miao2024spotserve}, and agentic workflow serving~\cite{peng2025hexgen}.

\begin{figure}
    \centering
    \includegraphics[width=0.5\linewidth]{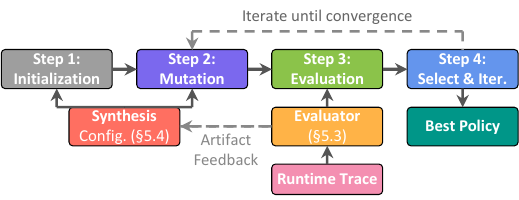}
    \caption{\small{Overview of LLM-driven program synthesis workflow. The evaluator (\S\ref{subsec:evaluator}) scores each candidate and produces artifact feedback. The synthesis configuration (\S\ref{subsec:synthesis-config}) configures the population initialization and guides mutations using the artifact feedback.}}
    \label{fig:evolve}
\end{figure}

\subsection{Workflow Overview}
\label{subsec:workflow}
The program synthesis workflow addresses the challenge of navigating this trade-off space by combining the code-generation capabilities of an LLM with population-based exploration to systematically explore the serving policy space. It proceeds as follows:
\begin{itemize}
    \item \textbf{Step 1: Initialization.} A population of candidate policies is initialized according to the synthesis configuration (\S\ref{subsec:synthesis-config}). Each candidate consists of a $\texttt{schedule}$ and $\texttt{should\_reschedule}$ function pair, representing a complete serving policy.
    \item \textbf{Step 2: Mutation.} In each iteration, the LLM mutates candidates from the current population to generate new policies, guided by the synthesis configuration (\S\ref{subsec:synthesis-config}).
    \item \textbf{Step 3: Evaluation and feedback.} The evaluator (\S\ref{subsec:evaluator}) executes the \texttt{should\_reschedule} and \texttt{schedule} functions of each candidate on the snapshotted runtime trace, producing both the fitness score $\text{T}_{\text{total}}$ and structured artifact feedback to guide subsequent mutations (\S\ref{subsec:synthesis-config}).
    \item \textbf{Step 4: Selection and iteration.} The synthesis workflow employs a MAP-Elites-inspired~\cite{mouret2015illuminating,sharma2025openevolve} candidate selection mechanism combined with island-based population management~\cite{tanese1989distributed} to retain high-performing candidates while maintaining population diversity. Steps 2-4 repeat until the evolutionary process converges or a computational budget is exhausted.
\end{itemize}
The illustration of this workflow is shown in~\autoref{fig:evolve}. This workflow operates within the control plane (\S\ref{sec:overview}), running asynchronously alongside live serving in the data plane.

\subsection{Evaluator}
\label{subsec:evaluator}

The evaluator is the critical bridge between the synthesis process and real-world runtime dynamics. Its role is to faithfully assess how a candidate policy would perform under the current serving conditions, capturing the deeply intertwined trade-offs that govern $\text{T}_{\text{total}}$. The evaluator achieves this through a trace-replay process backed by a serving simulator and produces structured artifact feedback to inform the mutation process.

\vspace{0.25em}
\noindent \textbf{Serving simulator.}
The serving simulator produces the cost estimates used by the evaluator during trace replay. It supports two types of estimation:
(\textbf{\underline{i}}) \textit{Serving time estimation} computes the expected serving time of a given plan under a given workload. It accepts four categories of input: The current workload characteristics (e.g., request arrival rates, sequence length distributions), model specifications (e.g., layer count, hidden size), hardware characteristics (e.g., compute TFLOPS, memory bandwidth, inter-GPU communication bandwidth), and a serving plan (i.e., workload assignment, resource allocation, and parallelism strategy). The output is used to derive both $\text{t}_{\text{serve}}$ and $\text{t}_{\text{stale}}$ in the execution model.
(\textbf{\underline{ii}}) \textit{Reconfiguration cost estimation} computes the transition overhead between two consecutive serving plans. It accepts four categories of input: Model specifications, transmission bandwidth, the previous plan, and the new plan. The output is used to derive $\text{t}_{\text{reconfig}}$ in the execution model. Internally, the simulator analytically models serving details including per-operator execution time~\cite{yuan2024llm}, continuous batching~\cite{yu2022orca}, model parallelism~\cite{shoeybi2019megatron}, and reconfiguration overhead~\cite{fu2024serverlessllm}. We detail the simulator's analytical methodology and validate its fidelity in~\autoref{appendix:serving-simu}.

\vspace{0.25em}
\noindent \textbf{Evaluator process.}
Rather than evaluating candidates \seqsplit{against} synthetic benchmarks, the evaluator replays the actual runtime trace snapshotted from the data plane (\S\ref{sec:overview}) against each candidate policy. The replay proceeds in three stages:
(\textbf{\underline{i}}) \textit{Monitoring and triggering:} The evaluator monitors the simulated workload at periodic time steps (e.g., 5 seconds per step). At each monitoring point, the candidate's $\texttt{should\_reschedule}$ function is invoked to determine whether rescheduling is warranted.
(\textbf{\underline{ii}}) \textit{Scheduling and simulation:} If rescheduling is triggered, the candidate's $\texttt{schedule}$ function produces a new plan, and the evaluator records its wall-clock scheduling time as $\text{t}_{\text{sched}}$. The evaluator then invokes the serving simulator to compute the remaining cost components for the triggered interval.
(\textbf{\underline{iii}}) \textit{Aggregation:} The per-interval costs across all triggered intervals are aggregated into $\text{T}_{\text{total}}$ as defined in Equation~\ref{eq:ttotal} (\S\ref{subsec:formulation}), ensuring that each candidate is scored against the runtime workload and cluster conditions observed during live serving.

\vspace{0.25em}
\noindent \textbf{Artifact feedback.}
Beyond the scalar fitness score $\text{T}_{\text{total}}$, the evaluator produces a structured cost breakdown as artifact feedback to the synthesis configuration, which drives the LLM-based population mutation (\textbf{Step 2}). As shown in~\autoref{tab:artifact}, for each evaluated candidate, the feedback reports the observed rescheduling frequency $\text{N}$ and the accumulated cost for each component ($\Sigma\text{t}_{\text{stale}}$, $\Sigma\text{t}_{\text{reconfig}}$, $\Sigma\text{t}_{\text{serve}}$). By comparing the cost breakdown of a child candidate against its parent, the LLM can identify how its code modification shifted the cost structure. For example, in~\autoref{tab:artifact}, Child~A is characterized by higher rescheduling frequency ($\text{N}$: $2 \to 5$), lower accumulated stale cost ($\Sigma\text{t}_{\text{stale}}$: $7.6\text{s} \to 5.5\text{s}$), and reduced reconfiguration cost ($\Sigma\text{t}_{\text{reconfig}}$: $12\text{s} \to 4\text{s}$), yielding a net improvement in $\text{T}_{\text{total}}$. In contrast, Child~B is characterized by lower rescheduling frequency ($\text{N}$: $2 \to 1$), higher accumulated stale cost ($\Sigma\text{t}_{\text{stale}}$: $7.6\text{s} \to 10\text{s}$), and increased reconfiguration cost ($\Sigma\text{t}_{\text{reconfig}}$: $12\text{s} \to 17\text{s}$), degrading $\text{T}_{\text{total}}$. From this, the LLM can infer that lightweight, frequent rescheduling outperforms thorough, infrequent replanning under the current workload, and steer mutations accordingly.

\begin{table}[t]
\centering
\caption{\small{Example artifact feedback comparing a parent candidate with two child candidates. The cost breakdown helps identify the strengths of high-performing candidates and informs the direction of subsequent code modifications.}}
\label{tab:artifact}
\small
\begin{tabular}{l|c|ccc|c}
\hline
\textbf{Candidate} & $\text{N}$ & $\Sigma\text{t}_{\text{stale}}$ & $\Sigma\text{t}_{\text{reconfig}}$ & $\Sigma\text{t}_{\text{serve}}$ & $\text{T}_{\text{total}}$ \\
\hline
Parent & 2 & 7.6s & 12.0s & 25.2s & 44.8s \\
Child A & 5 & 5.5s & 4.0s & 24.0s & 33.5s \\
Child B & 1 & 10.0s & 17.0s & 23.0s & 50.0s \\
\hline
\end{tabular}
\end{table}

\subsection{Synthesis Configuration}
\label{subsec:synthesis-config}
The synthesis configuration defines the operational parameters of the evolutionary process, and employs \textit{trade-off-aware mutation} to steer policy generation toward effective navigation of the runtime trade-off space.

\vspace{0.25em}
\noindent \textbf{Initialization.}
The synthesis configuration specifies the complete set of parameters governing the evolutionary process, including the evolutionary algorithm hyperparameters (e.g., maximum iterations, population size, number of islands) and LLM mutation settings (e.g., model identity, sampling weight, and temperature). The population is initialized with a small set of seed policies, each implementing the $\texttt{schedule}$ and $\texttt{should\_reschedule}$ function pair with a distinct algorithmic approach (e.g., greedy scheduling with periodic rescheduling, ILP-based optimization with threshold-based triggers), providing the LLM with a diverse starting vocabulary of design patterns from which to mutate. We provide a detailed initialization configuration in~\autoref{appendix:init-config}.
 
\vspace{0.25em}
\noindent \textbf{Trade-off-aware mutation.}
The key to effective policy synthesis lies in making each LLM mutation informed by the runtime trade-offs it must navigate. To this end, our trade-off-aware mutation applies diff-based code modifications~\cite{novikov2025alphaevolve} to high-performing candidates, integrating two complementary sources of guidance into the mutation process:
\begin{itemize}
    \item \textbf{Formal structure of runtime trade-offs.} This trade-off structure provides the LLM with a persistent understanding of the optimization landscape. The system prompt encodes (\textbf{\underline{i}}) the execution model structure (\autoref{eq:ttotal}) so the LLM understands how each cost component contributes to the end-to-end trace completion time, along with (\textbf{\underline{ii}}) concrete trade-off characterizations (\S\ref{sec:motivation}) so the LLM can calibrate the coupling between trade-off axes and reason about how a modification along one axis shifts the optimal point on the others (detailed system prompt example in~\autoref{appendix:system-prompt}).
    \item \textbf{Per-iteration dynamic feedback.} The mutation prompt is augmented at each iteration with dynamic information that grounds the LLM in the current evolutionary state. (\textbf{\underline{i}}) The artifact feedback from the evaluator (\S\ref{subsec:evaluator}) exposes the cost breakdown of both the parent and its children candidates, allowing the LLM to compare their cost structures, pinpoint which code modifications led to improvement or regression, and steer subsequent mutations accordingly. (\textbf{\underline{ii}}) The population context (obtained from \textbf{Step 4}) supplies the best-so-far fitness score and a summary of recently explored strategies, directing the LLM away from redundant mutations and toward underrepresented regions of the trade-off space.
\end{itemize}
By unifying these sources, each mutation is simultaneously informed by \textit{what} the trade-off landscape is and \textit{where} the current policy stands within it, enabling the LLM to produce trade-off-aware modifications instead of random variations.

\section{Implementation}
\label{sec:impl}
This section details the implementation of \sys's two-plane system architecture (mentioned in \S\ref{sec:overview}), focusing on the optimizations that enable the evolutionary process to operate efficiently (\S\ref{subsec:evolution-opt}) and the deployment mechanisms that integrate the self-evolving loop into live serving (\S\ref{subsec:deployment-impl}).

\subsection{Evolution Optimization}
\label{subsec:evolution-opt}

\noindent \textbf{Warm-start re-evolution.}
Consecutive runtime trace snapshots often reflect gradually shifting conditions, meaning the previously evolved policy is a strong starting point for the next evolution cycle. Rather than restarting from scratch, \sys implements a \textit{warm-start re-evolution} mechanism analogous to local-search-based rescheduling in online optimization~\cite{sun2024llumnix}: When a new snapshot triggers cycle $\text{e}_{\text{i}}$ (\autoref{fig:workflow}), the control plane initializes the population with the top-performing candidates from $\text{e}_{\text{i-1}}$ along with their mutations. This both accelerates convergence, as the LLM applies targeted refinements rather than re-exploring the trade-off space from scratch, and enables the system to consume more recent snapshots, keeping the deployed policy closely aligned with the latest runtime conditions.

\vspace{0.25em}
\noindent \textbf{Multi-level timeout hierarchy.}
LLM-generated candidate policies may contain expensive or divergent computation that would otherwise stall the evolutionary process. To maintain robustness, \sys employs a multi-level timeout hierarchy. An \textit{evolution-level timeout} bounds the total wall-clock time of each evolution cycle, ensuring the control plane delivers an updated policy within a predictable latency. A \textit{candidate-level timeout} bounds the execution time of each candidate's scheduling algorithm, terminating candidates whose $\texttt{schedule}$ function enters unbounded computation. This ensures degenerate candidates are discarded without stalling the evolutionary process.

\subsection{Deployment and Online Serving}
\label{subsec:deployment-impl}

\begin{figure}
    \centering
    \includegraphics[width=0.5\linewidth]{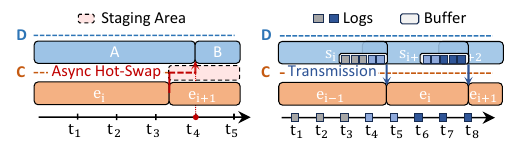}
    \caption{\small{Illustration of policy hot-swap (\underline{\textbf{left}}) and trace snapshotting (\underline{\textbf{right}}). D, C represent data and control planes; A, B represent different policies. $\text{s}_\text{i}$, $\text{e}_\text{i}$, and $\text{t}_\text{i}$ represent the $\text{i}$-th snapshot, evolution, and monitoring time point.}}
    \label{fig:two-impl}
\end{figure}

\noindent \textbf{Asynchronous policy hot-swap.}
The decoupled two-plane architecture requires a deployment mechanism that transfers an evolved policy from the control plane to the data plane without disrupting live serving. Consequently, \sys implements an \textit{asynchronous policy hot-swap} mechanism, as shown in~\autoref{fig:two-impl} (left). Since the serving policy is a pair of pure functions ($\texttt{schedule}$ and $\texttt{should\_reschedule}$), swapping the policy reduces to a lightweight code replacement with negligible overhead. When the control plane produces a superior policy, it writes the new code to a shared staging area, and the data plane asynchronously loads the replacement at its next monitoring time step. The new policy takes effect immediately: Subsequent invocations of $\texttt{should\_}\allowbreak\texttt{reschedule}$ and $\texttt{schedule}$ immediately execute the evolved code without disrupting any in-flight serving plan.

\vspace{0.25em}
\noindent \textbf{Runtime trace snapshotting.}
The data plane must efficiently capture runtime conditions and transmit them to the control plane without introducing overhead on live request processing. \sys implements a \textit{sliding-window snapshotting} mechanism that operates in two stages. In the \textit{recording stage}, the data plane continuously logs workload and cluster conditions at periodic monitoring time steps, appending each time point to a fixed-size circular buffer. In the \textit{transmission stage}, when the control plane signals for a new snapshot (i.e., upon completion of the current evolution process), the data plane extracts a contiguous window of recent time points from this buffer and transmits it via a dedicated background thread, decoupled from live request processing. The window size is configurable and determines the temporal horizon over which the control plane optimizes, and consecutive snapshots may overlap in their time-point coverage, allowing the control plane to observe workload trends that span across evolution cycles. The illustration of the mechanism is shown in~\autoref{fig:two-impl} (right).

\section{System Evaluation}
\label{sec:eva}


This evaluation answers three essential questions:
\begin{itemize}
     \item \textit{Q1: What is the end-to-end performance of Autopoiesis compared to state-of-the-art serving systems under diverse runtime dynamics?} (\S\ref{subsec:end-to-end})
     \item \textit{Q2: How does the self-evolving process reshape serving policies throughout deployment?}(\S\ref{subsec:policy_analysis})
     \item \textit{Q3: How effective is the LLM-driven program synthesis workflow in discovering high-performing policies?} (\S\ref{subsec:evolution})
 \end{itemize}


\noindent \textbf{Environments.}
We consider three cluster configurations: (\textbf{\underline{i}}) \textit{Homogeneous cluster:} 32 NVIDIA H100-80G GPUs across 4 nodes (8 GPUs/node), NVLink at 300\,GB/s intra-node, InfiniBand at 50\,GB/s inter-node. Workload fluctuation is the primary source of runtime dynamics. (\textbf{\underline{ii}}) \textit{Heterogeneous cluster:} 64 GPUs spanning four types (20$\times$A100-40G, 20$\times$A100-80G, 8$\times$H100-80G, 16$\times$H20-96G), NVLink at 300\,GB/s intra-node, 20\,Gb/s inter-node. Workload fluctuation under hardware heterogeneity is the primary source of runtime dynamics. (\textbf{\underline{iii}}) \textit{Elastic cloud cluster:} Up to 64 A100-80G GPUs across 8 nodes, with GPU availability fluctuating between 24 to 64 GPUs due to spot instance preemption and replenishment. NVLink at 300\,GB/s intra-node, 20\,Gb/s inter-node. Elastic cluster size is the primary source of runtime dynamics.

\vspace{0.25em}
\noindent \textbf{Baselines.}
Three state-of-the-art LLM serving systems are considered, each targeting one category of runtime dynamics:
(\textbf{\underline{i}}) DistServe~\cite{zhong2024distserve}, a disaggregated LLM serving framework that implements resource and parallelism scheduling to handle workload fluctuations, evaluated on the homogeneous cluster;
(\textbf{\underline{ii}}) HexGen~\cite{jiang2025demystifying}, a heterogeneous LLM serving framework that generates deployment plans to adapt to diverse workload and hardware heterogeneity, evaluated on the heterogeneous cluster;
(\textbf{\underline{iii}}) SpotServe~\cite{miao2024spotserve}, an elastic LLM serving framework that handles cluster dynamics through adaptive parallelism reconfiguration and stateful inference recovery, was evaluated on the elastic cloud cluster.

\vspace{0.25em}
\noindent \textbf{Traces.}
For the homogeneous and heterogeneous cluster evaluations, we use workload traces subsampled from \seqsplit{ShareGPT}~\cite{zheng2023judging} and LongBench~\cite{bai2024longbench}. For the elastic cloud cluster evaluation, we use cluster membership traces derived from the Microsoft Azure Functions (MAF) dataset~\cite{shahrad2020serverless}, combined with ShareGPT workload traces for request generation. All traces are augmented with synthetic workload fluctuation patterns, following standard practice in LLM serving evaluation~\cite{wang2025burstgpt,sun2024llumnix}, to stress-test policy adaptation under shifting runtime conditions (trace information in~\autoref{appendix:trace}).

\vspace{0.25em}
\noindent \textbf{Models and metrics.}
We deploy Llama3.1-70B as the served model on homogeneous and elastic cloud clusters, and additionally deploy Llama3.1-8B and Llama2-13B on the heterogeneous cluster. For evaluation, we report trace completion time, consisting of scheduling, reconfiguration, and serving time as defined in our execution model (\autoref{eq:ttotal}).

\vspace{0.25em}
\noindent \textbf{Synthesis LLM.}
The LLM-driven program synthesis workflow (\S\ref{sec:llm-driven}) is powered by GLM-4.7-Flash~\cite{zeng2025glm}. Our framework is model-agnostic and can be driven by any capable LLM.

\subsection{End-to-end Evaluation}
\label{subsec:end-to-end}

\begin{figure*}[t]
    \centering
    \includegraphics[width=\linewidth]{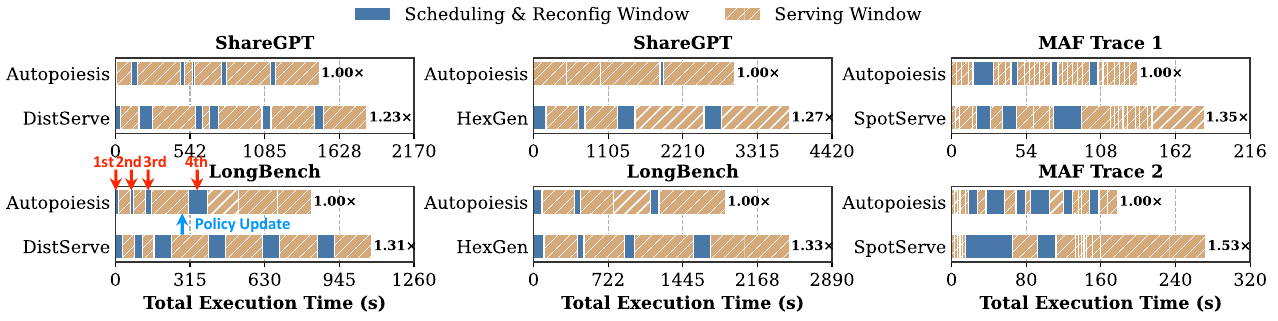}
    \caption{\small{End-to-end system execution time comparison between \sys and DistServe (\underline{\textbf{left}}), HexGen (\underline{\textbf{middle}}), SpotServe (\underline{\textbf{right}}) on homogeneous, heterogeneous, and elastic cloud clusters with different testing traces.}}
    \label{fig:distserve}
    \label{fig:hexgen}
    \label{fig:spotserve}
\end{figure*}


\noindent\textbf{Evaluation on homogeneous cluster with DistServe.}
\autoref{fig:distserve} (\textbf{\underline{left}}) compares \sys with DistServe on a homogeneous cluster under two representative traces. Overall, \sys achieves up to 31\% and on average 27\% reduction in total execution time.
During volatile workload periods that require frequent plan adaptation, \sys reduces scheduling and reconfiguration overhead by 65\% to rapidly respond to workload drifts (e.g., the 2nd and 3rd scheduling \& reconfiguration windows on the LongBench trace). Although per-period serving latency increases marginally by 2\%, the net effect is a 25\% reduction in end-to-end execution time for those periods.
Conversely, during stable workload periods, where conditions do not necessitate frequent rescheduling, \sys invests 10\% more time in scheduling and reconfiguration to search for a superior serving plan (e.g., the 4th scheduling \& reconfiguration window on the LongBench trace). This reduces serving latency by 22\% and the period's end-to-end execution time by 11\%. Notably, this behavioral shift occurs because the serving policy is updated before the 4th window to adapt to the stable regime.
Together, these results demonstrate that \sys is workload-adaptive: it minimizes reconfiguration overhead when workload volatility is high to maintain responsiveness, and invests more search effort when the workload is stable to discover superior plans. This adaptive behavior enables \sys to consistently outperform DistServe, which applies a fixed scheduling policy regardless of runtime dynamics.


\vspace{0.25em}
\noindent\textbf{Evaluation on heterogeneous cluster with HexGen.} 
As shown in ~\autoref{fig:hexgen} (\textbf{\underline{middle}}), \sys achieves up to 33\% and on average 30\% reduction in end-to-end execution time compared to HexGen on a heterogeneous cluster.
On heterogeneous clusters, the scheduling search becomes a critical bottleneck---HexGen requires over 300\,s to complete policy search in most scenarios due to the combinatorial explosion of device-heterogeneous parallelism configurations. To address this, \sys autonomously discovers scheduling principles tailored to device heterogeneity, reducing policy search time while minimizing the impact on serving latency. We present a detailed analysis of the evolved scheduling policies and their qualitative differences in~\S\ref{subsec:policy_analysis}.


\vspace{0.25em}
\noindent\textbf{Evaluation on elastic cloud cluster with SpotServe.} As shown in~\autoref{fig:spotserve} (\textbf{\underline{right}}), we compare \sys with SpotServe on an elastic cloud cluster. \sys achieves up to 53\% and on average 44\% reduction in total execution time.
Unlike previous scenarios, the dominant runtime trade-off in elastic cloud clusters is between reconfiguration time and serving performance. When the cluster composition remains stable, \sys invests in migrating to an optimal serving plan, accepting higher reconfiguration overhead to minimize serving latency. When frequent instance preemptions and arrivals occur, \sys instead migrates to an intermediate plan between the current and optimal configurations, reducing reconfiguration overhead at a marginal cost to serving performance. In contrast, SpotServe either reconfigures to the plan returned by its fixed search procedure or selects the configuration with the lowest monetary cost (i.e., using fewer instances), failing to exploit this runtime trade-off. These results further confirm that \sys consistently outperforms baselines that apply fixed scheduling policies regardless of runtime dynamics.

\subsection{Serving Policy Deep Dive}
\label{subsec:policy_analysis}

\vspace{0.25em}
\noindent\textbf{Case study}: \textit{Evolved serving policy in the heterogeneous case.} We analyze the qualitative differences between the evolved scheduling policy discovered by \sys and HexGen's original policy on heterogeneous clusters. 
The evolved policy minimizes scheduling time while maintaining near-optimal serving performance through several scenario-specific scheduling principles:
(\textbf{\underline{i}}) Bounding tensor parallelism communication within intra-machine to prune low-performance candidates (learned from evolution experience that inter-machine tensor parallelism introduces significant communication bottlenecks on heterogeneous interconnects);
(\textbf{\underline{ii}}) adjusting the workload assignment granularity from uniform multiples to curated sweet spots, reducing the number of decision variables;
(\textbf{\underline{iii}}) adding a constraint that every model must have at least one active serving group to reject invalid candidates early;
(\textbf{\underline{iv}}) making GPU assignment heterogeneity-aware by prioritizing the most suitable GPU type for each model, e.g., H100 for Llama-3.1-70B (learned from evolution experience that compute-intensive large models benefit disproportionately from high-performance GPUs, while smaller models tolerate weaker hardware with minimal throughput loss); and
(\textbf{\underline{v}}) setting a weighted secondary objective that prioritizes latency reduction for the bottleneck model (learned from evolution experience that overall serving throughput is dominated by the slowest model's latency in multi-model co-serving).
All modifications are discovered autonomously through \sys's LLM-driven program synthesis workflow. Note that all modifications reduce scheduling time by up to 92.2\% while reaching near-optimal serving efficiency.


\subsection{Effectiveness of the Evolution Process}
\label{subsec:evolution}

\begin{figure}
    \centering
    \includegraphics[width=0.5\linewidth]{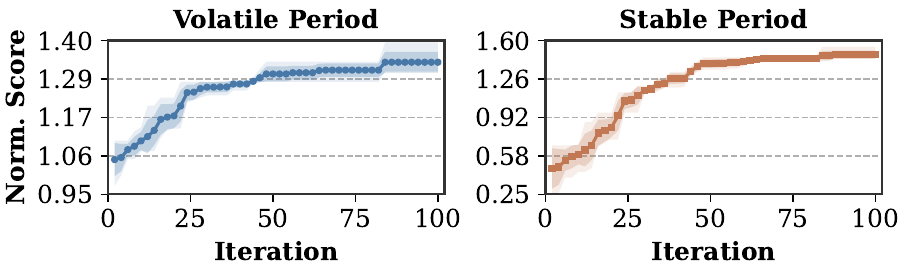}
    \caption{\small{Convergence curve of the evolution process on volatile and stable workload periods in the heterogeneous cluster. We run the evolution multiple times with different random seeds. Scores are normalized by the initial score.}}
    \label{fig:convergence-curve}
\end{figure}

\noindent\textbf{Convergence analysis.} As shown in~\autoref{fig:convergence-curve}, we evaluate the convergence behavior of \sys's evolution process under different workload characteristics in the heterogeneous cluster setup. Under volatile and stable workload periods, the evolution converges within approximately 80 and 60 iterations (less than 10 and 8 minutes), respectively. Both settings reach a stable plateau, confirming that the evolutionary process reliably discovers high-quality policies regardless of workload characteristics. We note that the evaluation stage of the evolutionary search is inherently parallelizable, as population candidates are evaluated independently; with 10 parallel evaluation threads, the evolution process can be accelerated by up to 8$\times$.

\begin{figure}
    \centering
    \includegraphics[width=0.5\linewidth]{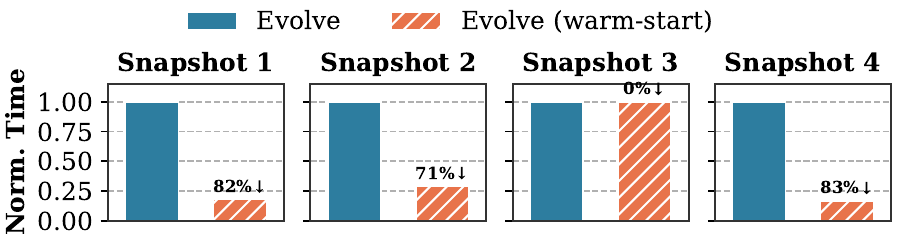}
    \caption{\small{Normalized evolution time across four different runtime snapshots, comparing cold-start evolution (Evolve) against warm-start re-evolution (Evolve warm-start).}}
    \label{fig:warm-start}
\end{figure}

\vspace{0.25em}
\noindent \textbf{Warm-start impact on evolution process.} As shown in~\autoref{fig:warm-start}, warm-start re-evolution effectively reduces evolution time in the majority of snapshots, by up to 83\% and on average 79\%. This confirms that when consecutive snapshots reflect gradually shifting conditions, previously evolved policies serve as strong initialization points, enabling the LLM to apply targeted refinements rather than re-exploring the trade-off space from scratch. However, when runtime conditions shift substantially such that the prior population offers no reusable structure, the warm-started search converges no faster than a cold start. Overall, warm-start re-evolution enables \sys to allocate its evolutionary budget more efficiently, avoiding redundant exploration when runtime conditions are relatively stable.

\section{System Analysis and Discussion}
\label{sec:analysis}
This section complements \S\ref{sec:eva} with controlled case studies. Each study isolates a specific workload regime (e.g., volatile or stable) and compares the evolved policy against static baselines at opposite extremes of the trade-off space (e.g., greedy vs. ILP), revealing how our LLM-driven synthesis workflow (\S\ref{sec:llm-driven}) navigates runtime trade-offs.

\subsection{Workload Fluctuation}
\label{subsec:workload-fluc}

\noindent \textbf{Setup.}
We consider an LLM serving deployment at the Swiss AI Center~\cite{yao2026opentela}, where a heterogeneous GPU cluster serves multiple LLMs under dynamic workloads. The serving policy must manage scheduling, reconfiguration, and serving trade-offs, and determine the rescheduling frequency to minimize overall system execution time as workload conditions shift.

\vspace{0.25em}
\noindent \textbf{Traces.} We subsample two workload traces from Swiss AI Center~\cite{yao2026opentela} with distinct fluctuation patterns: A \textit{volatile trace} with frequently changing characteristics and a \textit{stable trace} with seldom changing characteristics. Detailed scheduling problem formulation and workload trace information are shown in~\autoref{appendix:case-study} and \autoref{appendix:trace}.

\vspace{0.25em}
\noindent \textbf{Baselines.} We compare our evolved policy against two static baselines that represent opposite extremes of the trade-off space: A \textit{greedy-based policy} optimized for scheduling speed, and an \textit{ILP-based policy} optimized for plan quality, both with manually tuned rescheduling frequencies.

\begin{table}[t]
\centering
\caption{\small{End-to-end comparison under workload fluctuation.}}
\label{tab:workload-fluctuation}
\small
\resizebox{0.5\linewidth}{!}{%
\begin{tabular}{l|l|c|ccc|c}
\hline
\textbf{Trace} & \textbf{Policy} & \textbf{N} & $\Sigma\text{t}_{\text{stale}}$ & $\Sigma\text{t}_{\text{reconfig}}$ & $\Sigma\text{t}_{\text{serve}}$ & \text{Overall Thpt} \\
\hline
\multirow{3}{*}{Volatile}
& Greedy & 10 & 65.1s & 244.8s & 422.1s & 95,723 t/s \\
& ILP & 1 & 104.7s & 0.0s & 686.5s & 88,558 t/s \\
& \textbf{Ours} & \textbf{9} & \textbf{17.5s} & \textbf{217.6s} & \textbf{416.3s} & \textbf{107,578 t/s} \\
\hline
\multirow{3}{*}{Stable}
& Greedy & 10 & 27.2s & 38.0s & 120.5s & 436,353 t/s \\
& ILP & 1 & 34.3s & 0.0s & 113.3s & 548,851 t/s \\
& \textbf{Ours} & \textbf{2} & \textbf{4.3s} & \textbf{4.2s} & \textbf{115.8s} & \textbf{651,222 t/s} \\
\hline
\end{tabular}%
}
\end{table}

\vspace{0.25em}
\noindent \textbf{Results analysis.}
\autoref{tab:workload-fluctuation} summarizes the results. Averaged across both traces, the greedy-based policy reschedules at every transition (N=10), achieving comparable serving latency to the evolved policy (271.3s vs. 266.1s) but accumulating 1.5$\times$ higher combined scheduling and reconfiguration overhead (187.6s vs. 121.8s, composed of 46.1s scheduling + 141.4s reconfiguration), resulting in a 1.2$\times$ degradation in $\text{T}_{\text{total}}$ compared to the evolved policy (458.8s vs. 387.8s). The ILP-based policy reschedules only once at initialization (N=1), reducing combined overhead to 69.5s but serving under a stale plan as workload conditions shift, producing 50\% higher serving latency than the evolved policy (399.9s vs. 266.1s) and a 1.2$\times$ degradation in $\text{T}_{\text{total}}$ (469.4s vs. 387.8s). Notably, the greedy-based policy outperforms the ILP-based policy on the volatile trace (732.0s vs. 791.2s) where frequent rescheduling matters most, while the ILP-based policy outperforms the greedy-based policy on the stable trace (147.6s vs. 185.6s) where the initial plan remains effective, indicating that neither baseline is universally superior. The evolved policy reduces combined overhead to 121.8s (10.9s scheduling + 110.9s reconfiguration) while maintaining the lowest serving latency, adapting to different workload traces by balancing the scheduling and reconfiguration overhead against the serving latency to minimize the overall end-to-end time.

\vspace{0.5em}
\begin{mdframed}[style=insightbox]
\small
\textbf{\underline{Insight 1:} Evolved policies reshape the Pareto frontier.} The scheduling algorithm design space exhibits a Pareto frontier between two competing costs: Scheduling and reconfiguration ($\Sigma\text{t}_{\text{stale}} + \Sigma\text{t}_{\text{reconfig}}$) vs.\ serving time ($\Sigma\text{t}_{\text{serve}}$). More thorough scheduling and reconfiguration typically result in lower serving time. Conventional policies (e.g., greedy, ILP) occupy fixed points on this frontier, trading one cost for the other. LLM-driven synthesis escapes this fixed frontier through two complementary behaviors: (\textbf{\underline{i}}) \textit{Frontier-shifting.} The evolved policy discovers workload-specific scheduling principles that simultaneously improve both scheduling and reconfiguration cost and serving time, reaching operating points no conventional policy accesses. (\textbf{\underline{ii}}) \textit{Frontier-navigating.} Where simultaneous improvement on both axes is not possible, the evolved policy locates the narrow efficient region on the existing frontier, accepting a marginal concession on one axis in exchange for the majority of achievable savings on the other. This dual behavior demonstrates that the synthesis workflow tailors not only policy parameters but the fundamental algorithmic strategy to the current workload structure.
\end{mdframed}
\vspace{0.5em}
 
\vspace{0.25em}
\noindent \textbf{Rescheduling strategy analysis.}
The greedy-based policy, due to its minimized per-rescheduling overhead, chooses to reschedule at every transition across both traces (N=10), outperforming the ILP-based policy when the workload is volatile but accumulating unnecessary scheduling and reconfiguration overhead on stable intervals. In contrast, the ILP-based policy, due to its high scheduling cost, reschedules only once at initialization (N=1), outperforming the greedy-based policy when the workload is stable but suffering from stale plans during volatile intervals. The evolved rescheduling strategy adapts its frequency to the workload's volatility: Under the volatile trace, it reschedules 9 times to keep plans fresh; under the stable trace, it reschedules only twice, avoiding overhead with minimal latency impact.

\vspace{0.5em}
\begin{mdframed}[style=insightbox]
\small
\textbf{\underline{Insight 2:} Scheduling algorithm and rescheduling strategy require co-evolution.} Rescheduling frequency and scheduling algorithm design are fundamentally coupled: A lightweight scheduler can afford frequent rescheduling but yields diminishing plan quality, while a computationally intensive scheduler produces superior plans but cannot sustain the overhead of frequent invocation. Conventional approaches are constrained by this coupling because their algorithmic cost profiles are fixed at design time. Co-evolution breaks this coupling by reshaping the algorithm's computational profile to match the rescheduling frequency each workload regime demands, rather than forcing the frequency to accommodate a fixed overhead.
\end{mdframed}
\vspace{0.5em}

\subsection{Elastic Cluster Dynamics}
\label{subsec:spotinstance}

\noindent \textbf{Setup.}
We consider the same deployment setting as \S\ref{subsec:workload-fluc}, but where the cluster topology changes during serving, driven by two common sources: Involuntary changes from spot instance preemption~\cite{miao2024spotserve} (e.g., AWS \texttt{g4dn} instances being reclaimed) and voluntary changes from autoscaling mechanisms~\cite{fu2024serverlessllm} (e.g., scaling out under increased load or scaling in during low utilization). The serving objective remains consistent with \S\ref{subsec:workload-fluc}. In both cases, rescheduling is mandatory at each cluster transition, as the previous plan may reference unavailable hardware or leave newly added resources idle. The key trade-off therefore shifts from scheduling thoroughness and frequency to \textit{reconfiguration aggressiveness}: How much of the serving plan to reconfigure at each transition. 

\vspace{0.25em}
\noindent \textbf{Traces.} We subsample two traces from the Microsoft Azure Functions (MAF) dataset~\cite{shahrad2020serverless} with volatile and stable cluster change characteristics (scheduling problem formulation and trace information in~\autoref{appendix:case-study} and \autoref{appendix:trace}).

\vspace{0.25em}
\noindent \textbf{Baselines.} We compare against two static baselines: A \textit{full-migration policy} optimized for serving performance by always reconfiguring to the globally optimal plan, and a \textit{minimal-migration policy} optimized for reconfiguration efficiency by always reconfiguring to the nearest operational plan.

\begin{table}[t]
\centering
\caption{\small{E2E comparison under elastic cluster dynamics.}}
\label{tab:elastic-cluster}
\small
\resizebox{0.5\linewidth}{!}{%
\begin{tabular}{l|l|cc|c}
\hline
\textbf{Trace} & \textbf{Policy} & $\Sigma\text{t}_{\text{stale}}$ & $\Sigma\text{t}_{\text{reconfig}}$ & $\text{Overall Thpt}$ \\
\hline
\multirow{3}{*}{Volatile}
& Full-migration & 9.7 & 108.8s &  8,433 t/s \\
& Minimal-migration & 1.6s & 8.4s &  13,304 t/s \\
& \textbf{Ours} & \textbf{0.7s} & \textbf{44.4s} & \textbf{14,843 t/s} \\
\hline
\multirow{3}{*}{Stable}
& Full-migration & 9.6s & 67.9s & 11,080 t/s \\
& Minimal-migration & 2.1s & 6.3s & 10,187 t/s \\
& \textbf{Ours} & \textbf{0.6s} & \textbf{30.8s} & \textbf{12,731 t/s} \\
\hline
\end{tabular}%
}
\end{table}

\vspace{0.25em}
\noindent \textbf{Results analysis.}
\autoref{tab:elastic-cluster} summarizes the results. The elastic cluster dynamics scenario is primarily governed by the trade-off between reconfiguration aggressiveness and serving performance. The full-migration policy achieves the highest plan quality on both traces at the cost of heavy scheduling and reconfiguration overhead (14.9s $+$ 108.8s on volatile, 9.6s $+$ 67.9s on stable), while the minimal-migration policy minimizes overhead (2.4s $+$ 6.3s on volatile, 2.1s $+$ 6.3s on stable) at the cost of degraded serving throughput. As a result, minimal-migration prevails on the volatile trace where reconfiguration overhead dominates (13,304 t/s vs. 8,433 t/s), while full-migration prevails on the stable trace where plan quality dominates (11,080 t/s vs. 10,187 t/s). The evolved policy achieves the highest throughput on both traces (14,843 t/s and 12,731 t/s) while incurring moderate reconfiguration overhead (30.8s), improving over the better-performing baseline by 16\% on the volatile trace and 15\% on the stable trace. This suggests that the evolved policy discovers reconfiguration strategies that capture the majority of the serving benefit without full plan replacement.

\vspace{0.5em}
\begin{mdframed}[style=insightbox]
\small
\textbf{\underline{Insight 3:} Reconfiguration strategy similarly benefits from evolution.} In elastic cluster dynamics scenarios where reconfiguration aggressiveness becomes the primary trade-off, an optimal reconfiguration strategy must (\textbf{\underline{i}}) navigate the trade-off between reconfiguration aggressiveness and serving performance, and (\textbf{\underline{ii}}) find the minimum reconfiguration necessary to achieve near-optimal serving quality. The evolved policy addresses both by incorporating and fine-tuning reconfiguration-aware mechanisms (e.g., per-model migration cost-benefit checks, reconfiguration penalty magnitude adjustment), simultaneously controlling the overall aggressiveness and minimizing unnecessary transitions.
\end{mdframed}
\vspace{0.5em}

\subsection{Agentic Request Scheduling}
\label{subsec:req-scheduling}
 
\noindent \textbf{Setup.}
We consider a request scheduling problem in LLM serving for agentic workflows, where the system must determine request-to-GPU assignment and queuing order within each GPU's local queue before each round of request processing. Unlike the previous scenarios that operate at cluster-level deployment granularity, scheduling decisions here are made round by round, and the objective is to minimize end-to-end request processing time. The core trade-off is therefore direct: Scheduling computation time competes with request completion time.

\vspace{0.25em}
\noindent \textbf{Trace.} We use two non-overlapping 64-workflow slices subsampled from ShareGPT~\cite{zheng2023judging}. Each workflow is a multi-call agentic request trace with online call revelation. Detailed scheduling problem formulation and trace information are provided in~\autoref{appendix:case-study} and~\autoref{appendix:trace}.

\vspace{0.25em}
\noindent \textbf{Baselines.} We compare against two static baselines: An \textit{MILP-based policy} that solves the request assignment and queuing problem optimally but incurs high per-round scheduling overhead, and a \textit{greedy-based policy} that produces assignments quickly but struggles to achieve optimal request completion time. \autoref{tab:req-scheduling} summarizes the results.

\vspace{0.25em}
\noindent \textbf{Execution model calibration.} As this scenario operates at the request level rather than the cluster level, no model migration or parallelism reconfiguration occurs between rounds. The execution model (\S\ref{subsec:formulation}) naturally accommodates this by eliminating the reconfiguration component, reducing $\text{T}_{\text{total}}$ to the sum of per-round scheduling overhead and serving time (i.e., $\Sigma\text{t}_{\text{sched}}+\Sigma\text{t}_{\text{serve}}$).

\begin{table}[t]
\centering
\caption{End-to-end comparison under asynchronous request scheduling on two ShareGPT traces.}
\label{tab:req-scheduling}
\small
\begin{tabular}{l|ccc|ccc}
\hline
& \multicolumn{3}{c|}{\textbf{Trace 1}} & \multicolumn{3}{c}{\textbf{Trace 2}} \\
\textbf{Policy} & $\Sigma\text{t}_{\text{sched}}$ & $\Sigma\text{t}_{\text{serve}}$ & $\text{T}_{\text{total}}$ & $\Sigma\text{t}_{\text{sched}}$ & $\Sigma\text{t}_{\text{serve}}$ & $\text{T}_{\text{total}}$ \\
\hline
Greedy & 0.1s & 10.4s & 10.5s & 0.1s & 18.0s & 18.1s \\
MILP & 6.9s & 25.0s & 31.9s & 2.2s & 10.3s & 12.5s \\
\textbf{Ours} & \textbf{1.3s} & \textbf{5.6s} & \textbf{6.8s} & \textbf{0.4s} & \textbf{6.7s} & \textbf{7.1s} \\
\hline
\end{tabular}
\end{table}

 
\vspace{0.25em}
\noindent \textbf{Results analysis.}
The evolved policy outperforms all baselines on both traces. On Trace~1, the greedy-based policies maintain negligible scheduling overhead but produce weaker serving plans, yielding $\text{T}_{\text{total}}$ of $10.5\text{s}$. The MILP-based policy optimizes each round exactly but becomes scheduling-dominated: Its accumulated scheduling overhead reaches $6.9\text{s}$, inflating $\text{T}_{\text{total}}$ to $31.9\text{s}$. In contrast, the evolved policy reduces scheduling time to $1.3\text{s}$ while simultaneously improving serving quality, achieving a $\text{T}_{\text{total}}$ of $6.8\text{s}$ (35\% reduction over the greedy baseline and 79\% over the MILP-based policy). A similar trend holds on the harder Trace~2, where the evolved policy achieves $\text{T}_{\text{total}}$ of $7.1\text{s}$, outperforming the greedy-based ($18.1\text{s}$, 61\% reduction) and MILP-based ($12.5\text{s}$, 43\% reduction) policies.

\vspace{0.5em}
\begin{mdframed}[style=insightbox]
\small
\textbf{\underline{Insight 4:} Our LLM-driven program synthesis workflow adapts across serving scenarios.} The LLM-driven program synthesis workflow (\S\ref{subsec:evaluator}) adapts to different types of runtime dynamics (from cluster-level deployment to request-level scheduling) by tailoring the execution model (\S\ref{subsec:formulation}) to the system cost structure of each scenario. In request scheduling, the evolved policy synthesizes domain-specific principles (e.g., priority-aware queuing, load-aware affinity) that approximate optimal solutions at a fraction of the computational cost, demonstrating our workflow's ability to discover efficient operating points in diverse trade-off spaces and serving scenarios.
\end{mdframed}
\vspace{0.5em}

\vspace{0.25em}
\noindent \textbf{Additional case studies and insights.} We also present one additional case studies in~\autoref{appendix:additional-case-study}. It isolates the scheduling algorithm as the optimization target, revealing that the synthesis workflow discovers deployment-tailored algorithmic optimizations.

\vspace{0.25em}
\noindent \textbf{Discussion.} While \sys demonstrates strong performance across diverse runtime dynamics, several practical considerations remain regarding deployment cost, LLM reliability, configuration sensitivity, and evolutionary efficiency. We provide a detailed discussion in~\autoref{appendix:discussion}.

\label{sec:conclusion}
\section{Conclusion}
We present \sys, a self-evolving serving paradigm that shifts LLM serving from static, human-engineered policy design to continuous, LLM-driven policy evolution. By decoupling policy execution from policy evolution through a two-plane architecture, \sys continuously observes runtime dynamics, evolves serving policies via LLM-driven program synthesis, and deploys superior policies without disrupting live serving. Evaluations across different serving scenarios demonstrate up to 53\% and on average 34\% improvement over state-of-the-art serving systems. We believe \sys establishes a new paradigm in which serving policies are no longer fixed artifacts but living code that evolves alongside the runtime conditions.

\bibliographystyle{unsrt}
\bibliography{main}

\clearpage

\appendix

\begin{table*}[h!t]
\centering
\caption{Notations.}
\label{tab:notation}
\resizebox{\linewidth}{!}{%
\begin{tabular}{l|l}
\hline
\textbf{Symbol} & \textbf{Description} \\
\hline
$i \in \{1, \ldots, N\}$ & Timestamp index; $N$ is the total number of timestamps. \\
$\sigma_i$ & Scheduling plan at timestamp $i$, specifying model assignment and parallelism. \\
$\boldsymbol{\sigma} = \{\sigma_1, \dots, \sigma_N\}$ & Sequence of scheduling plans over the time horizon. \\
$\mathbf{T}_{\text{total}}$ & Total end-to-end latency over the time horizon. \\
\hline
$d \in \mathbf{D}$ & GPU device in the heterogeneous device set $\mathbf{D} = \{d_1, \ldots, d_K\}$. \\
$m_d$ & Memory capacity of device $d$. \\
$g \in \mathbf{G}$ & GPU type in the type set $\mathbf{G} = \{g_1, \ldots, g_P\}$. \\
$\kappa_g$ & Total device count for GPU type $g$. \\
$m_g$, $F_g$, $B_g$, $P_g$ & Per-device HBM capacity, peak FP16 throughput, HBM bandwidth, PCIe bandwidth of type $g$. \\
\hline
$\mathbf{M} = \{M_1, \ldots, M_Z\}$ & Set of $Z$ models. \\
$\mathbf{W}_z = \{ w_{z,1}, \ldots, w_{z,J_z} \}$ & Set of $J_z$ workload types for model $M_z$. \\
$W_z$ & Total weight size of model $M_z$ in bytes (see \autoref{eq:weight-mem}). \\
$\lambda_{z,i}$ & Total concurrent request count (batch size) for $M_z$ at timestamp $i$. \\
$s^p_{z,i}$, $s^d_{z,i}$ & Prefill (input) and decode (output) lengths for $M_z$ at timestamp $i$. \\
$b^{\max}_{z,i}$ & Maximum per-replica batch size for $M_z$ at timestamp $i$. \\
\hline
$\mathcal{T}_g$ & Set of valid tensor-parallel (TP) degrees for GPU type $g$. \\
$\mathcal{B}_{z,i}$ & Candidate per-replica batch sizes for $M_z$ at timestamp $i$. \\
$n_{z,g,t,b} \in \mathbb{Z}_{\ge 0}$ & Number of replicas of $M_z$ on GPU type $g$ with TP degree $t$ and batch size $b$. \\
$y_{z,g,t,b} \in \{0,1\}$ & Indicator: 1 iff replica group $(z,g,t,b)$ is active. \\
$L_z$ & Completion latency of model $M_z$ (slowest active replica group). \\
$T_{\text{balanced}}$ & Global makespan: $\max_{z \in \mathbf{M}} L_z$. \\
\hline
$\textsc{Serve-Cost}(\sigma_i)$ & Serving latency to process workloads under plan $\sigma_i$. \\
$\textsc{Reconfig-Cost}(\sigma_{i-1}, \sigma_i)$ & Reconfiguration overhead between consecutive scheduling plans. \\
$\textsc{Stale-Cost}(\sigma_i)$ & Penalty for stale or delayed serving at timestamp $i$. \\
$\textsc{Sched-Cost}$ & Scheduling overhead (ILP solve time). \\
$\tau(z, g)$ & Weight transfer time for model $M_z$ on GPU type $g$ via PCIe. \\
$c_z$ & PCIe bandwidth efficiency coefficient for model $M_z$. \\
\hline
\end{tabular}%
}
\end{table*}
\section{Notations}

We summarize the notations used in the paper in \autoref{tab:notation}.

\section{Serving Simulator}
\label{appendix:serving-simu}

We present the concrete cost models used in our roofline-based simulator to instantiate $\textsc{Serve-Cost}$, $\textsc{Reconfig-Cost}$, and $\mathbf{T}_{\text{total}}$. Each GPU type $g \in \mathbf{G}$ has per-device HBM capacity $m_g$, peak FP16 throughput $F_g$ (FLOPs/s), HBM bandwidth $B_g$ (bytes/s), PCIe bandwidth $P_g$ (bytes/s), and $\nu_g$ GPUs per physical node. Each model $M_z$ is described by: layers $L_z$, hidden dimension $H_z$, MLP intermediate dimension $I_z$, vocabulary size $V_z$, attention heads $A_z$, key-value heads $K_z \leq A_z$, per-head dimension $d_z = H_z / A_z$, and weight precision $\eta$ bits.

\noindent \textbf{Model Weight Size.} The total parameter footprint of $M_z$ in bytes is:

\begin{small}
\begin{equation}
  \label{eq:weight-mem}
  W_z
  = \Bigl[
      L_z \bigl(
        3 H_z I_z
        + 2 A_z H_z d_z
        + 2 K_z H_z d_z
      \bigr)
      + 2 H_z V_z
    \Bigr]
    \cdot \tfrac{\eta}{8}
\end{equation}
\end{small}
where the per-layer terms account for the three MLP projections ($3H_z I_z$), the query and output attention projections ($2 A_z H_z d_z = 2H_z^2$), the key and value projections ($2 K_z H_z d_z$), and the shared embedding and output-projection matrices ($2 H_z V_z$).

\noindent \textbf{Roofline-Based Serve Cost.}
The latency $\Lambda(z,g,t,b,s^p,s^d)$ for a replica group serving $M_z$ on GPU type $g$ with TP degree $t$ and per-replica batch size $b$ is estimated via the hardware roofline model. For an operation requiring $\mathcal{F}$ FLOPs and accessing $\mathcal{A}$ bytes, the arithmetic intensity is $\mathrm{AI} = \mathcal{F}/\mathcal{A}$ and the ridge point is $\mathrm{AI}^* = F_g / B_g$. The achieved performance and operation time are:
\begin{small}
\begin{equation}
\begin{aligned}
  \label{eq:roofline}
  \pi(\mathcal{F}, \mathcal{A}, g)
  = \min\!\left(\frac{\mathcal{F}}{\mathcal{A}} \cdot B_g,\; F_g\right)
\end{aligned}
\end{equation}
\end{small}

\begin{small}
\begin{equation}
\begin{aligned}
  \label{eq:roofline-2}
  T^{\mathrm{op}}(\mathcal{F}, \mathcal{A}, g)
  = \frac{\mathcal{F}}{\pi(\mathcal{F}, \mathcal{A}, g)}
\end{aligned}
\end{equation}
\end{small}

The total serve latency sums $T^{\mathrm{op}}$ across all transformer sublayers for both the prefill and decode phases, plus the TP all-reduce communication overhead:
\begin{small}
\begin{equation}
\label{eq:serve-latency}
\begin{aligned}
\Lambda(z,g,t,b,s^p,s^d)
&= T^{\mathrm{prefill}}(z,g,t,b,s^p) \\
&+ \sum_{k=s^p}^{s^p + s^d - 1}
    T^{\mathrm{decode}}(z,g,t,b,k)
\end{aligned}
\end{equation}
\end{small}
where $T^{\mathrm{prefill}}$ and $T^{\mathrm{decode}}$ apply \autoref{eq:roofline} and \autoref{eq:roofline-2} to each sublayer and add the ring-allreduce time incurred at each of the $L_z$ transformer layers (two all-reduces per layer: after the attention output projection and after the MLP down-projection):
\begin{small}
\begin{equation}
  \label{eq:allreduce}
  T^{\mathrm{comm}}(z,g,t,b,s)
  = \frac{2(t-1)}{t} \cdot 2 L_z \cdot H_z \cdot b \cdot s
    \cdot \tfrac{\eta}{8} \cdot R_{g,t}^{-1}
\end{equation}
\end{small}
where $s \in \{1, s^p\}$ is the sequence length for the decode and prefill phases respectively, and $R_{g,t}$ is the effective interconnect bandwidth (intra-node when $t \leq \nu_g$, cross-node otherwise).

\noindent\textbf{Memory Feasibility.}
Under TP degree $t$, each GPU shard of $M_z$ holds $W_z / t$ bytes of model weights. A group $(z,g,t,b)$ is \textit{infeasible} if the per-shard weight exceeds available device capacity, i.e., OOM issues occur:
\begin{small}
\begin{equation}
  \label{eq:oom}
  W_z / t \;>\; \theta \cdot m_g
\end{equation}
\end{small}
where $\theta = 0.8$ is a memory utilization threshold reserving headroom for the KV cache. Infeasible groups are assigned a penalty latency $\Lambda_\infty \gg 0$; the min-makespan objective then naturally avoids activating them without requiring additional constraints.

\noindent \textbf{Reconfiguration Cost.} When consecutive plans $\sigma_{i-1}$ and $\sigma_i$ differ for model $M_z$ — i.e., the $(\text{GPU type},\,\text{TP},\,\text{count})$ tuples changes — the cluster must terminate old replicas and reload new ones via PCIe. Let $\Delta\mathbf{M}_i \subseteq \mathbf{M}$ denote the set of models whose placement changed from $\sigma_{i-1}$ to $\sigma_i$. The reconfiguration cost decomposes as:
\begin{small}
\begin{equation}
  \label{eq:reconfig}
  \textsc{Reconfig-Cost}(\sigma_{i-1}, \sigma_i)
  = T^{\mathrm{term}}(\sigma_{i-1}, \sigma_i)
  + T^{\mathrm{load}}(\sigma_i)
\end{equation}
\end{small}

\noindent For model $M_z$ on GPU type $g$, the weight transfer time is TP-insensitive (the full model weight is transferred regardless of sharding degree):
\begin{small}
\begin{equation}
  \label{eq:transfer-time}
  \tau(z, g) = \frac{W_z}{P_g} \cdot c_z
\end{equation}
\end{small}
where $c_z \in [5.3,\, 11.5]$ is a model-size-dependent PCIe bandwidth utilization coefficient (smaller models incur higher per-byte overhead due to PCIe protocol amortization).

\noindent The \textit{termination cost} is the time to drain and evict all changed models from HBM; since different GPU types can terminate concurrently, the bottleneck is the slowest group:
\begin{small}
\begin{equation}
  \label{eq:term-cost}
  T^{\mathrm{term}}(\sigma_{i-1}, \sigma_i)
  = \max_{z \in \Delta\mathbf{M}_i}\;
    \max_{\substack{g \in \mathbf{G} :\\ \sigma_{i-1}\text{ deploys }M_z\text{ on }g}}
    \tau(z, g)
\end{equation}
\end{small}

\noindent The \textit{loading cost} is the time to transfer new model weights to all target GPU types; since different GPU types load in parallel, the bottleneck is again the slowest:
\begin{small}
\begin{equation}
  \label{eq:load-cost}
  T^{\mathrm{load}}(\sigma_i)
  = \max_{g \in \mathbf{G}}\;
    \max_{\substack{z \in \mathbf{M} :\\ \sigma_i\text{ deploys }M_z\text{ on }g}}
    \tau(z, g)
\end{equation}
\end{small}
If $\sigma_{i-1} = \sigma_i$, both costs are zero.

\section{Scheduling Problem Formulation}
\label{appendix:case-study}

\subsection{Formulation for Case Studies}

\noindent\textbf{Scheduling problem formulation.} Given a heterogeneous cluster of GPUs, we consider the problem of jointly scheduling multiple models under time-varying workloads. Let $\mathbf{D} = \{d_1, \ldots, d_K\}$ denote a set of $K$ heterogeneous GPU devices, where each device $d$ has a memory capacity $m_d$ and distinct compute characteristics. Let $\mathbf{M} = \{M_1, \ldots, M_Z\}$ denote a set of $Z$ models. Each model $M_z$ may serve multiple types of workloads (e.g., inference requests with different input sizes or other requirements), denoted as $\mathbf{W}_z = \{w_{z,1}, \ldots, w_{z,J_z}\}$.

We consider a discrete time horizon of $N$ timestamps. At each timestamp $i \in \{1, \ldots, N\}$, the system observes a workload demand $\mathcal{W}_i$ across all models and workload types. A scheduling plan at timestamp $i$ is denoted as $\sigma_i$, which specifies (\textbf{\underline{i}}) the assignment of models and workloads to devices, and (\textbf{\underline{ii}}) the parallel execution configuration, i.e., data, pipeline, or tensor parallelism, for each model. In our context, the data parallelism means the number of inference replicas. 

The scheduling plan must satisfy device memory constraints:
\vspace{-0.25em}
\begin{small}
\begin{equation}
\textsc{Mem-Cumsum}(d, i) \leq m_d, \quad \forall d \in \mathbf{D}, \; \forall i \in \{1, \ldots, N\}
\end{equation}
\end{small}

\vspace{-0.25em}
\noindent where $\textsc{Mem-Cumsum}(d, i)$ denotes the total memory consumption on device $d$ at timestamp $i$.

Each scheduling plan $\sigma_i$ incurs four types of costs:
\begin{itemize}
    \item $\textsc{Serve-Cost}(\sigma_i)$: the serving latency (or execution time) to process workloads at timestamp $i$;
    \item $\textsc{Reconfig-Cost}(\sigma_{i-1}, \sigma_i)$: the reconfiguration overhead due to changes in scheduling decisions across consecutive timestamps, i.e., from $\sigma_{i-1}$ to $\sigma_i$;
    \item $\textsc{Stale-Cost}(\sigma_i)$: the penalty incurred due to stale or delayed serving at timestamp $i$;
    \item $\textsc{Sched-Cost}$: the scheduling overhead at the first timestamp.
\end{itemize}

The end-to-end latency for serving the workload over the time horizon is:
\vspace{-0.25em}
\begin{small}
\begin{equation}
\label{eq:ttotal}
\begin{aligned}
\mathbf{T}_{\text{total}} = \;& \textsc{Sched-Cost} + \textsc{Serve-Cost}(\sigma_1) \\
& + \sum_{i=2}^{N} \Big[ \textsc{Stale-Cost}(\sigma_i) \\
\qquad &+ \textsc{Reconfig-Cost}(\sigma_{i-1}, \sigma_i) \\
 \qquad &+ \textsc{Serve-Cost}(\sigma_i) \Big]
\end{aligned}
\end{equation}
\end{small}

Our objective is to find a sequence of scheduling plans $\boldsymbol{\sigma} = \{\sigma_1, \sigma_2, \ldots, \sigma_N\}$ that minimizes the total trace completion time:
\vspace{-0.25em}
\begin{small}
\begin{equation}
\label{eq:objective}
\begin{aligned}
\min_{\boldsymbol{\sigma}} \quad & \mathbf{T}_{\text{total}} \\
\text{s.t.} \quad 
& \textsc{Mem-Cumsum}(d, i) \leq m_d, \quad \\
& \forall d \in \mathbf{D}, \; \forall i \in \{1, \ldots, N\}
\end{aligned}
\end{equation}
\end{small}

\vspace{-0.25em}
\noindent The search space of possible scheduling plans is combinatorial due to (\textbf{\underline{i}}) heterogeneous device capabilities, (\textbf{\underline{ii}}) multiple models and workload types, (\textbf{\underline{iii}}) flexible parallel execution strategies, and (\textbf{\underline{iv}}) temporal dependencies introduced by reconfiguration costs. Consequently, identifying the optimal sequence of scheduling plans is NP-hard. In practice, we seek efficient heuristic or approximate algorithms to find a near-optimal solution.

\subsection{Formulation for Agentic Request}
\noindent\textbf{Agentic Request Scheduling Formulation.}
Given a disaggregated prefill-decode serving cluster, we consider the problem of online scheduling for multi-call agentic workflows. Let $\mathbf{P}$ and $\mathbf{Q}$ denote the sets of prefill and decode instances, respectively. Each decode instance has a finite token-capacity budget, and different instances may have different serving characteristics.

Let $\mathbf{\Omega}=\{\omega_1,\ldots,\omega_B\}$ denote a set of workflow requests. Each workflow $\omega$ arrives over time and involves a series of LLM calls, each with an input length and an output length. A call becomes schedulable only after all of its predecessor calls have completed, so the scheduler must make decisions under online call revelation.

At each scheduling trigger $i \in \{1,\ldots,N\}$, the system observes the currently waiting calls and computes a scheduling plan $\sigma_i$. A plan specifies (\textbf{\underline{i}}) the assignment of each waiting call to a prefill instance and a decode instance, and (\textbf{\underline{ii}}) the queue order on each instance. The plan must satisfy stage-wise capacity constraints: prefill instances process one call at a time, while decode instances must respect cumulative token-capacity limits.

Each workflow $\omega$ has a committed horizon $SLO_\omega$ and an actual completion time $C_\omega$. The quality of a scheduling policy is measured by the completion ratio $C_\omega/SLO_\omega$, and we use trace-level metrics such as $95^{th}$ percentile, $99^{th}$ percentile, and mean $C_\omega/SLO_\omega$ to evaluate end-to-end performance.

Because scheduling is performed online, each plan $\sigma_i$ incurs both a scheduling overhead and a serving cost. Let $\textsc{Sched-Cost}(\sigma_i)$ denote the planning time at trigger $i$, and let $\textsc{Serve-Cost}(\sigma_i)$ denote the resulting serving latency contribution. The total trace completion cost is:
\vspace{-0.25em}
\begin{small}
\begin{equation}
\mathbf{T}_{\text{total}}
=
\sum_{i=1}^{N}
\Big[
\textsc{Sched-Cost}(\sigma_i)
+
\textsc{Serve-Cost}(\sigma_i)
\Big]
\end{equation}
\end{small}

Our objective is to find a sequence of scheduling plans
$\boldsymbol{\sigma}=\{\sigma_1,\sigma_2,\ldots,\sigma_N\}$
that minimizes end-to-end workflow completion degradation, i.e., improves high-percentile and average $C_\omega/SLO_\omega$ while keeping scheduling overhead low enough for online use. The search space is combinatorial due to heterogeneous resources, multi-stage assignment and ordering decisions, token-capacity constraints, and online workflow revelation. Consequently, exact online optimization is expensive, and in practice we seek efficient schedulers that achieve a strong quality-latency trade-off.

\section{Initialization Configuration}
\label{appendix:init-config}

In Table \ref{tab:init}, we present the initialization configuration of \sys for the evolutionary process. These configurations are fine-tuned to achieve the optimal performance in the experiments presented in our paper.

\begin{table}[h]
\centering
\vspace{-1em}
\caption{Configuration parameters used in \sys during the evolutionary process.}
\label{tab:init}
\small
\resizebox{0.5\linewidth}{!}{%
\begin{tabular}{c|c|c}
\hline
\textbf{Configuration Scope} & \textbf{Parameter Name} & \textbf{Configured Value} \\
\hline
\multirow{8}{*}{LLM} 
 & Primary Model & Claude Opus 4.6 \\ 
 & Sampling Weights & 0.7 \\ \cline{2-3}
 & Secondary Model & Claude Sonnet 4.6 \\
 & Sampling Weights & 0.3 \\ \cline{2-3}
 & Temperature & 0.7 \\ \cline{2-3}
 & Max Tokens & 16384 \\ \cline{2-3}
 & Max Code Length & 50000 \\ \hline
\multirow{4}{*}{Evolutionary}
 & Max Iterations & 100 \\ \cline{2-3}
 & Population Size & 50 \\ \cline{2-3}
 & Number of Islands & 3 \\ \cline{2-3}
 & Elite Selection Ratio & 0.2 \\ \hline
\end{tabular}%
}
\end{table}

\section{System Prompts}
\label{appendix:system-prompt}

In this section, we present examples of system prompts we used in \sys. Across evolutionary iterations, the following prompts are used to guide the LLM to fine-tune the generated program in the next evolutionary iteration.

\noindent \textbf{Search–serving trade-off guidance:} For a given workload and heterogeneous cluster, fine-grained exploration of the candidate search space generally yields higher-quality scheduling decisions and lower serving latency. In contrast, coarse-grained exploration tends to produce suboptimal scheduling outcomes and increased serving latency. Accordingly, prompts should be carefully designed to guide the LLM in balancing search granularity against scheduling quality. Our detailed prompt message for this guidance is shown in \underline{prompt message episode 1}.

\noindent \textbf{Reconfiguration and serving trade-off guidance:} More comprehensive reconfiguration generally results in lower serving latency but incurs higher reconfiguration costs. Conversely, less extensive reconfiguration tends to reduce reconfiguration costs at the expense of increased serving latency. Therefore, we guide the LLM to balance the trade-off between reconfiguration cost and serving latency. The detailed prompt message is presented in \underline{prompt message episode 2}.

\vspace{0.5em}
\begin{mdframed}[style=codebox]
\small
\textbf{\underline{Prompt Message Episode 1:} } 

\noindent\textbf{Search Granularity vs. Scheduling Efficiency}

\texttt{max\_batch\_per\_replica} controls how finely the ILP searches over batch sizes for each model. It represents a fundamental \textbf{search granularity vs. scheduling efficiency}
trade-off — there is no universally correct value; the right choice depends on the model's memory footprint and workload characteristics.

\textbf{Larger cap $\rightarrow$ fine-grained search, lower scheduling efficiency:}
\begin{itemize}
    \item More batch-size candidates $\rightarrow$ larger ILP $\rightarrow$ slower solve time.
    \item The scheduler can find placements that match the workload's \texttt{total\_batch} more precisely (e.g., a replica batch of 96 vs.\ only 64), potentially reducing latency.
    \item Each extra candidate adds integer and binary variables, making the solver (e.g., CBC) work harder.
\end{itemize}

\textbf{Smaller cap $\rightarrow$ coarse-grained search, higher scheduling efficiency:}
\begin{itemize}
    \item Fewer candidates $\rightarrow$ smaller ILP $\rightarrow$ faster solve time.
    \item The scheduler selects from a coarser set of batch sizes and may slightly over- or under-shoot the ideal per-replica load.
    \item For models where latency is relatively insensitive to batch size in the relevant range, the quality loss is negligible while the speed gain is significant.
\end{itemize}

\textbf{Key Insight: Some models need fine-grained search, others do not.}
\begin{itemize}
    \item Models with large \texttt{total\_batch} and flat latency-vs-batch curves benefit from finer granularity, enabling better packing.
    \item Memory-constrained models have few feasible batch sizes regardless of the cap — increasing it introduces infeasible (OOM) candidates, adding overhead without benefit.
    \item Models with steep latency-vs-batch curves benefit from a moderate cap that captures the efficient region without exploring far into OOM.
\end{itemize}

\textbf{Additional Insight:} Raising the cap for small models can reduce the number of required replicas, thereby freeing GPUs for larger models.  
For example, if a small model has \texttt{total\_batch = 256} and the cap is set to 256, only one replica is required (versus 8 replicas at cap = 32), freeing 7 GPUs that can be allocated to larger models to reduce latency.

\textbf{Implementation Guidance:} The \textbf{initial program} uses a uniform cap (defined in \texttt{workloads.py}, which is fixed), providing a safe and conservative baseline. LLM-evolved programs should instead choose per-model caps within their EVOLVE-BLOCKS based on model characteristics (e.g., weight size, memory footprint, and typical batch sizes). These caps can be different across timestamps. Inspect \texttt{workloads.py} to understand the \texttt{total\_batch} range for each model across timestamps.
\end{mdframed}

\vspace{0.5em}
\begin{mdframed}[style=codebox]
\small
\noindent\textbf{\underline{Prompt Message Episode 2:}}

\textbf{The Reconfiguration Trade-off — When to Reschedule}

The total cost for one timestamp is:
\begin{equation}
\text{e2e} = \text{ilp solve time} + \text{serving latency} + \text{reconfig cost}
\end{equation}

A new schedule that saves $\Delta$ seconds of serving and scheduling latency is only worthwhile if:
\begin{equation}
\Delta > \text{reconfig cost}
\end{equation}

\textbf{Concrete Trade-off Examples}

\textbf{Example A — Reschedule pays off:}

Suppose at $ts_0$ you have a large model on fast GPUs with serving latency $2000$ ms.  
At $ts_1$, traffic shifts, and you can reduce serving latency to $800$ ms by moving
to different GPUs (the following values serve as examples, and you should refer to the simulators for accurate values):
\begin{itemize}
    \item Loading cost: $\sim 0.25$ s (high TP, fast PCIe).
    \item Termination cost: $\sim 0.25$ s (high TP from previous config).
    \item $\text{reconfig cost} = 0.25 + 0.25 = 0.50$ s.
    \item Serving latency gain: $(2000 - 800)$ ms $= 1200$ ms $= 1.2$ s.
    \item Net gain: $1.2 - 0.5 = +0.7$ s saved $\rightarrow$. \textbf{RESCHEDULE} (profitable)
\end{itemize}

\textbf{Example B — Reschedule does NOT pay off:}

At $ts_1$, you can improve serving latency from $2500$ ms to $2100$ ms (saving $400$ ms $= 0.4$ s)
by moving a large model to a different GPU type/TP configuration (the following values serve as examples, and you should refer to the simulators for accurate values):
\begin{itemize}
    \item Loading cost: $\sim 0.51$ s.
    \item Termination cost: $\sim 0.25$ s.
    \item $\text{reconfig cost} = 0.51 + 0.25 = 0.76$ s.
    \item Serving latency gain: $0.4$ s.
    \item Net gain: $0.4 - 0.76 = -0.36$ s $\rightarrow$.
    \textbf{STAY} (reconfiguration costs more than it saves).
\end{itemize}

\textbf{Example C — Small-model reconfiguration is cheap:}

Moving a small model (13GB) from one GPU type to another (the following values serve as examples, and you should refer to the simulators for accurate values):
\begin{itemize}
    \item Loading cost: $\frac{13}{1 \times 64} = 0.20$ s.
    \item Termination cost: $\frac{13}{1 \times 32} = 0.41$ s.
    \item $\text{reconfig cost} = 0.20 + 0.41 = 0.61$ s.
\end{itemize}
If this frees GPUs for larger models and saves $2$ s of latency,  
net gain $= +1.39$ s $\rightarrow$ \textbf{profitable}.

\textbf{Example D — Staying put is sometimes optimal:}

If $ts_0 \rightarrow ts_1$ traffic change is mild (same models, $10$--$20\%$ batch size variation), the optimal plan barely differs. In this case:
\begin{itemize}
    \item Expected latency gain from rescheduling: $\sim 100$ ms $= 0.1$ s.
    \item $\text{reconfig cost}$ for large model: $\sim 0.5$ s.
    \item Net: $-0.4$ s $\rightarrow$ \textbf{STAY PUT} (keep the same placement from $ts_0$ to $ts_1$).
\end{itemize}
\end{mdframed}

\section{Additional Case Study}
\label{appendix:additional-case-study}

\subsection{Scheduling Algorithm Optimization}
\label{subsec:sched-opt}

\noindent \textbf{Setup.}
We consider a real-world LLM serving deployment at the Swiss AI Center~\cite{yao2026opentela}, where a heterogeneous GPU cluster serves multiple LLMs under dynamic workloads. In this scenario, we isolate the scheduling algorithm as the optimization target: Given a fixed workload snapshot and cluster configuration, the objective is to evolve a scheduling algorithm that produces serving plans (i.e., workload assignments, resource allocations, and parallelism strategies) of equivalent quality to a ILP-based~\cite{schrijver1998theory} scheduling algorithm while minimizing plan generation time.

\vspace{0.25em}
\noindent \textbf{Traces.}
We use two workload snapshots from Swiss AI Center~\cite{yao2026opentela}, each representing a distinct workload and cluster configuration. Detailed problem formulation and trace information are shown in~\autoref{appendix:case-study} and \autoref{appendix:trace}.

\vspace{0.25em}
\noindent \textbf{Results analysis.}
The baseline ILP-based scheduler requires an average of 120s per invocation across the two traces. The evolved scheduling algorithm reduces this to 9s, a $13{\times}$ speedup, with less than 3\% degradation in plan quality. Examining the evolved code reveals complementary modifications across multiple algorithmic layers: Search space reduction through curated candidate selection, tighter LP relaxation via per-variable bounds, valid inequalities encoding domain-specific constraints, strengthened secondary objectives for better solution guidance, and optimized solver parameters. A detailed comparison of the baseline and evolved code is provided in~\autoref{appendix:evolved-sched}.

\vspace{0.5em}
\begin{mdframed}[style=insightbox]
\small
\textbf{\underline{Insight 5:} Synthesis discovers deployment-tailored algorithmic optimizations.} Rather than converging on a single type of improvement, the synthesis workflow discovers complementary optimizations spanning problem formulation, domain knowledge injection, objective shaping, and solver configuration. Crucially, these optimizations are tailored to the specific workload and cluster configuration and would shift under different deployment conditions, highlighting the value of LLM-driven program synthesis over manual algorithm tuning.
\end{mdframed}
\vspace{0.5em}

\section{Evolved Scheduling Algorithm Analysis}
\label{appendix:evolved-sched}

In this section, we analyze the details of the code generated by LLMs. The LLM-driven improvements include smarter batch selection, per-variable bounds, stronger weighted objective, and solver-specific tuning. These enhancements significantly reduce ILP solve time while preserving the optimal makespan, highlighting the value of integrating domain knowledge with automated code refinement. The details are discussed below.

\noindent \textbf{Optimization on batch candidate selection.}  
In the LLM-generated program, the selection of batch candidates was refined compared to the naive exhaustive search employed in the initial program, which served as the starting point for the LLM's evolutionary improvements. Instead of traversing all possible batch sizes, the LLM generates a curated set of candidates, combining powers of 2, divisors of the total batch size, and manually selected sweet spots. Specifically, the resulting batch candidates are designed to balance computational efficiency and hardware utilization. The detailed batch candidates are shown as follows:

\begin{small}
\begin{equation}
\begin{aligned}
\text{Batch Candidates} &= \{  1, 2, 3, 4, 6 \} \;\cup\; \{ 2^k \mid k = 2, \dots, 6 \} \;\\&\cup\;  \{ 20, 24, 28, 32, 40, 48 \}
\end{aligned}
\end{equation}
\end{small}

\noindent \textbf{Optimization on parallel strategy.}  
The LLM incorporates a domain-specific constraint for large models (e.g., 70B), requiring a sufficiently high tensor parallelism degree, specifically enforcing $t \ge 4$. This ensures that large models are allocated adequate parallel resources, preventing inefficient configurations that often result in out-of-memory errors.

\noindent \textbf{Optimizations on the objective function and formulation.}  
To enhance the efficiency of the ILP solver, the LLM-generated code incorporates two key optimizations: (\textbf{\underline{i}}) tighter \textit{per-variable Big-M bounds}, which strengthen the LP relaxation, and (\textbf{\underline{ii}}) a \textit{weighted objective}, which provides improved guidance when multiple solutions share the same makespan. The notations used below are defined in \autoref{tab:notation}.

\begin{itemize}
\item \textbf{Per-variable Big-M bounds.}  
For model type $z$, GPU type $g$, tensor-paralel degree $t$, and batch size $b$, the standard indicator constraint linking $n_{z,g,t,b}$ and $y_{z,g,t,b}$ is: $n_{z,g,t,b} \le M \cdot y_{z,g,t,b}$, where $M$ is a sufficiently large constant; a naive formulation uses a single global bound $M = \max_{g \in \mathbf{G}} \kappa_g$, which is overly loose for most variables.

\noindent Instead, the LLM defines a \textit{variable-specific} bound for each possible tuple $(z,g,t,b)$:
\begin{small}
\begin{equation}
\begin{aligned}
M_{z,g,t,b} 
&= \min \Bigg( 
\left\lfloor \frac{\kappa_g}{t} \right\rfloor, \left\lceil \frac{\lambda_z}{b} \right\rceil 
\Bigg)
\end{aligned}
\end{equation}
\end{small}

\noindent The indicator constraint becomes:
\begin{small}
\begin{equation}
n_{z,g,t,b} \le M_{z,g,t,b} \cdot y_{z,g,t,b}
\end{equation}
\end{small}

\noindent Here, $\left\lfloor \frac{\kappa_g}{t} \right\rfloor$ bounds the number of replicas by available GPU capacity, $\left\lceil \frac{\lambda_z}{b} \right\rceil$ bounds the number of replicas required to cover the workload. The term \textit{per-variable} emphasizes that each decision variable $n_{z,g,t,b}$ has its own tight upper bound, instead of sharing a single global constant. This significantly tightens the LP relaxation: when $y_{z,g,t,b}$ takes fractional values, the feasible range of $n_{z,g,t,b}$ is correspondingly restricted. As a result, the solver obtains stronger bounds and can prune the branch-and-bound search tree more effectively.

\item \textbf{Weighted objective.} The primary objective is to minimize the global makespan:
\begin{small}
\begin{equation}
\min \; T_{\text{balanced}}, \quad 
\text{s.t. } T_{\text{balanced}} \ge L_z, \quad \forall z \in \mathbf{M}
\end{equation}
\end{small}

\noindent The variable $L_z$ is defined through:
\begin{small}
\begin{equation}
L_z \ge \Lambda_{z,g,t,b} \cdot y_{z,g,t,b}, \quad \forall g \in \mathbf{G}, \; t \in \mathcal{T}_g, \; b \in \mathcal{B}_z
\end{equation}
\end{small}

\noindent $L_z$ captures the completion latency of model $M_z$, determined by its slowest active replica group. While minimizing $T_{\text{balanced}}$ ensures fairness across models, multiple solutions may achieve the same optimal makespan but differ in individual model latencies. To guide the solver toward better practical solutions, the LLM introduces a weighted objective ($z$ indexes models in increasing order of size, i.e., a larger $z$ corresponds to a larger LLM):
\begin{small}
\begin{equation}
\text{Weighted Obj} = \epsilon \sum_{z \in \mathbf{M}} \big( 1.0 + 0.5 \cdot z \big) \, L_z,
\quad \epsilon = 0.05
\end{equation}
\end{small}

\begin{table}[ht]
\centering
\caption{Tuned CBC solver parameters for ILP acceleration}
\label{tab:cbc}
\resizebox{0.5\linewidth}{!}{%
\begin{tabular}{l | c | l}
\hline
\textbf{Parameter} & \textbf{Value} & \textbf{Description} \\
\hline
\multirow{3}{*}{\textsc{threads}} & \multirow{3}{*}{4} & Enable parallel solving across multiple \\
& & CPU threads, leveraging multiple \\
& & CPU cores for faster computation. \\ \hline
\multirow{2}{*}{\textsc{ratioGap}} & \multirow{2}{*}{0.003} & Stop the solver when the solution \\
& & is within 0.3\% of the optimal value.\\ \hline
\multirow{3}{*}{\textsc{strongBranching}} & \multirow{3}{*}{10} & Tuned branching parameter to guide \\
& & variable selection during \\
& & branch-and-bound. \\ \hline
\multirow{3}{*}{\textsc{passPresolve}} & \multirow{3}{*}{10} & Number of presolve passes to simplify \\
& & constraints and variables \\
& & before solving. \\ \hline
\multirow{3}{*}{\textsc{passCuts}} & \multirow{3}{*}{10} & Number of cutting passes to generate \\
& & additional valid inequalities \\
& & for tighter bounds. \\
\hline
\end{tabular}%
}
\end{table}

\noindent This strategy assigns larger weights to larger LLMs, encouraging the solver to reduce their latency when possible. Compared to an unweighted and very small secondary term, this stronger formulation provides a clearer optimization signal, helping the solver distinguish between solutions with identical makespan and converge faster to high-quality solutions.
\end{itemize}

\noindent \textbf{Optimization on ILP solver.} 
The ILP solver, i.e., CBC solver, was carefully tuned by the LLM. We summarize the details in \autoref{tab:cbc}. These settings directly accelerate ILP convergence while maintaining solution quality within a negligible 0.3\% optimality gap.

\section{Trace Information}
\label{appendix:trace}

\subsection{Trace Information for the Motivation Section}

We present two motivating traces using Qwen2.5 models (1.5B–72B). In the \textit{heavy-dominant} phase (H), small models (1.5B/3B/7B) serve low batch sizes (64/64/64) with 256-token prefill sequence length and short decode sequence lengths (2048/1536/3072 tokens), while large models (14B/32B/72B) carry high batch loads (384/256/128) with 512-token prefill length and long decode sequence lengths (8192/6144/5120 tokens); in the \textit{light-dominant} phase (L), small models dominate (960/480/288 batch, 256-token prefill sequence length, 4096/3072/6144-token decoding sequence length), while large models are minimally loaded (64/32/16 batch, 256-token prefill sequence length, 2048/1536/1280-token decoding sequence length).

\noindent \textbf{Trace for \autoref{fig:motivation} (left).} As shown in \autoref{tab:motivation-shifting}, this 3-timestamp trace features two abrupt phase transitions, demonstrating the scheduler's need for rapid response. The system starts in a heavy-dominant phase (ts0), abruptly shifts to a light-dominant phase (ts1), and reverts to heavy-dominant at ts2. Each transition dramatically redistributes load across all model sizes.

\begin{table}[t]
\centering
\caption{%
  \textit{Trace for \autoref{fig:motivation} (left)}: Two abrupt phase transitions across 3 timestamps (H$\to$L$\to$H). Phase transitions are marked with $\Rightarrow$.
}
\label{tab:motivation-shifting}
\resizebox{0.4\linewidth}{!}{%
\begin{tabular}{c|c|rrrrrr}
\hline
\textbf{TS} & \textbf{Phase}
  & \textbf{1.5B} & \textbf{3B} & \textbf{7B}
  & \textbf{14B} & \textbf{32B} & \textbf{72B} \\
\hline
ts0 & H               &  64 &  64 &  64 & 384 & 256 & 128 \\
ts1 & $\Rightarrow$ L & 960 & 480 & 288 &  64 &  32 &  16 \\
ts2 & $\Rightarrow$ H &  64 &  64 &  64 & 384 & 256 & 128 \\
\hline
\end{tabular}%
}
\end{table}

\noindent \textbf{Trace for \autoref{fig:motivation} (right).} As shown in \autoref{tab:motivation-hybrid}, this 5-timestamp trace hybridizes stable and volatile behavior. The workload begins in a light-dominant phase with minor perturbations (ts0–ts1), undergoes one abrupt shift to the heavy-dominant phase at ts2, and then stabilizes with minor perturbations (ts3–ts4). This demonstrates that even predominantly stable workloads benefit from adaptive rescheduling at shift boundaries.

\begin{table}[t]
\centering
\caption{%
  \textit{Trace for \autoref{fig:motivation} (right)}: 5 timestamps combining a stable period with one major phase shift (L$\to$H). Phase transition is marked with $\Rightarrow$.
}
\label{tab:motivation-hybrid}
\resizebox{0.4\linewidth}{!}{%
\begin{tabular}{c|c|rrrrrr}
\hline
\textbf{TS} & \textbf{Phase}
  & \textbf{1.5B} & \textbf{3B} & \textbf{7B}
  & \textbf{14B} & \textbf{32B} & \textbf{72B} \\
\hline
ts0 & L               & 960           & 480          & 288          &  64           &  32 &  16 \\
ts1 & L               & 968  & 476 & 284 &  64           &  32 &  16 \\
\hline
ts2 & $\Rightarrow$ H &  64           &  64          &  64          & 384           & 256 & 128 \\
\hline
ts3 & H               & 72   &  64          &  64          & 400  & 256 & 128 \\
ts4 & H               &  64           &  64          &  64          & 384           & 256 & 128 \\
\hline
\end{tabular}%
}
\end{table}

\subsection{Trace Information for the Case Studies}

\noindent \textbf{Stable workload in \S\ref{subsec:workload-fluc}.}
We consider a production workload consisting of 10 timestamps (ts0–ts9) and three Qwen2.5 model sizes (1.5B, 3B, and 7B), where batch sizes and sequence lengths remain mostly stable with occasional variations. The prefill length is fixed at 256 tokens for all models throughout. The decoding lengths are 4096, 3072, and 6144 tokens for the 1.5B, 3B, and 7B models, respectively. The slight variation exmpales include: at ts1 and ts8, the 1.5B batch increases by 5\% (960$\to$1008); at ts2, the 7B batch slightly decreases (288$\to$264); at ts3, the 1.5B \textit{decode} length doubles to 8192 tokens (batch unchanged); at ts4, the 3B batch increases by 13\% (480$\to$544); at ts6, the 7B batch increases by 17\% (288$\to$336) and its \textit{prefill} length doubles to 512 tokens. More detailed batch size variations are shown in \autoref{tab:workload-light}.

\begin{table}[t]
\centering
\caption{%
  \textit{Stable workload}: Batch sizes across 10 timestamps for three Qwen2.5 small models.
}
\label{tab:workload-light}
\resizebox{0.5\linewidth}{!}{%
\begin{tabular}{c|cccccccccc}
\hline
\textbf{Model} 
  & \textbf{ts0} & \textbf{ts1} & \textbf{ts2} & \textbf{ts3} & \textbf{ts4}
  & \textbf{ts5} & \textbf{ts6} & \textbf{ts7} & \textbf{ts8} & \textbf{ts9} \\
\hline
1.5B & 960 & 1008 & 952 & 960 & 968 & 956 & 962 & 958 & 1008 & 964 \\
3B   & 480 & 476  & 484 & 480 & 544 & 478 & 482 & 479 & 481  & 483 \\
7B   & 288 & 284  & 264 & 290 & 286 & 288 & 336 & 287 & 285  & 291 \\
\hline
\end{tabular}%
}
\end{table}

\noindent \textbf{Volatile workload in \S\ref{subsec:workload-fluc}.}
We use a highly dynamic workload consisting of 10 timestamps (ts0–ts9) and six Qwen2.5 model sizes (1.5B–72B), where the system alternates between two contrasting traffic regimes: a \textit{heavy-dominant} phase and a \textit{light-dominant} phase.
In the heavy phase, small models (1.5B/3B/7B) serve low batch sizes (64/64/64) with 256-token prefill length and short decode lengths (2048/1536/3072 tokens), while large models (14B/32B/72B) carry high batch loads (384/256/128) with 512-token prefill sequence length and long decode sequence lengths (8192/6144/5120 tokens).
In the light phase, the roles reverse: small models dominate with large batches (960/480/288), 256-token prefill sequence length, and decode sequence lengths of 4096/3072/6144 tokens, while large models are minimally loaded (64/32/16 batch) with 256-token prefill sequence length and decode sequence lengths of 2048/1536/1280 tokens. Detailed batch configurations are presented in \autoref{tab:workload-shifting-b}. Phase transitions at ts3, ts6, and ts9 encounter abrupt workload shifts.

\begin{table}[t]
\centering
\caption{%
\textit{Volatile workload}: Batch sizes per timestamp for six Qwen2.5 models. Phase transitions are marked with $\Rightarrow$.
}
\label{tab:workload-shifting-b}
\resizebox{0.4\linewidth}{!}{%
\begin{tabular}{c|c|rrrrrr}
\hline
\textbf{TS} & \textbf{Phase}
  & \textbf{1.5B} & \textbf{3B} & \textbf{7B}
  & \textbf{14B} & \textbf{32B} & \textbf{72B} \\
\hline
ts0 & H                  &  64           &  64 &  64           & 384           & 256 & 128 \\
ts1 & H                  & 80   &  64 &  64           & 400  & 256 & 128 \\
ts2 & H                  &  64           &  64 &  64           & 384           & 256 & 128 \\
\hline
ts3 & $\Rightarrow$ L    & 960           & 480 & 288           &  64           &  32 &  16 \\
ts4 & L                  & 1008 & 480 & 336  &  64           &  32 &  16 \\
ts5 & L                  & 960           & 480 & 288           &  64           &  32 &  16 \\
\hline
ts6 & $\Rightarrow$ H    & 96   &  64 &  64           & 416  & 256 & 128 \\
ts7 & H                  &  64           &  64 &  64           & 384           & 256 & 128 \\
ts8 & H                  & 80   &  64 &  64           & 400  & 256 & 128 \\
\hline
ts9 & $\Rightarrow$ L    & 960           & 480 & 288           &  64           &  32 &  16 \\
\hline
\end{tabular}%
}
\end{table}

\noindent \textbf{Cluster and workload information in \S\ref{subsec:spotinstance}.}
The cluster evolves over five timestamps (ts0--ts4) across three GPU types: H100-SXM (80\,GB, NVLink 900\,GB/s, PCIe 64\,GB/s), H200-SXM (141\,GB, NVLink 900\,GB/s, PCIe 64\,GB/s), and A100-80GB (80\,GB, NVLink 600\,GB/s, PCIe 32\,GB/s). We present the detailed stable cluster information in \autoref{tab:stable-clusters}, and the volatile cluster information in \autoref{tab:volatile-clusters}. The workload consists of three Qwen2.5 models (7B, 14B, and 72B). Requests for all models have a 512-token prefill sequence length. The decoding length for the 7B model is 512-token and the batch size is 32; the decoding length for the 14B model is 2048-token and the batch size is 64; the decoding length for the 72B model is 4096-token and the batch size is 128.

\begin{table}[t]
\centering
\caption{%
  \textit{Stable cluster transition in \S\ref{subsec:spotinstance}}: GPU counts per type across five timestamps. Cluster transitions are marked with $\Rightarrow$.
}
\label{tab:stable-clusters}
\resizebox{0.65\linewidth}{!}{%
\begin{tabular}{c|c|ccc|c}
\hline
\textbf{TS} & \textbf{Transition}
  & \textbf{A100-80GB} & \textbf{H100-SXM} & \textbf{H200-SXM}
  & \textbf{Total GPUs} \\
\hline
ts0 &  H100+H200                  &  --            & 16             & 16             & 32 \\
ts1 & $\Rightarrow$ H200 increased        &  --            & 16             & 24    & 40 \\
ts2 & $\Rightarrow$  H100 increased        &  --            & 24    & 24             & 48 \\
\hline
ts3 & $\Rightarrow$ A100 introduced, H100 and H200 reduced       & 16    & 16    & 8    & 40 \\
ts4 & $\Rightarrow$ A100 reduced, H100 and H200 increased & 8     & 24    & 16             & 48 \\
\hline
\end{tabular}%
}
\end{table}

\begin{table}[t]
\centering
\caption{%
  \textit{Volatile cluster transition in \S\ref{subsec:spotinstance}}: GPU counts per type across five timestamps. Cluster transitions are marked with $\Rightarrow$.
}
\label{tab:volatile-clusters}
\resizebox{0.7\linewidth}{!}{%
\begin{tabular}{c|c|ccc|c}
\hline
\textbf{TS} & \textbf{Transition}
  & \textbf{A100-80GB} & \textbf{H100-SXM} & \textbf{H200-SXM}
  & \textbf{Total GPUs} \\
\hline
ts0 & A100+H100+H200                  & 8 & 16 & 16 & 40 \\
ts1 & $\Rightarrow$ A100 and H100 reduced, H200 increased        & -- & 8  & 24 & 32 \\
ts2 & $\Rightarrow$ A100 introduced, H100 increased, H200 redcued        & 16 & 24 & 8  & 48 \\
\hline
ts3 & $\Rightarrow$ H100 spiked       & 16 & 40 & 8  & 64 \\
ts4 & $\Rightarrow$ A100, H100 reduced, H200 increased & 8  & 24 & 16 & 48 \\
\hline
\end{tabular}%
}
\end{table}

\subsection{Trace Information for the Main Experiments}
\noindent Across the main experiments, we use phase labels to indicate the dominant token-shape characteristic of each workload segment. A \textit{prefill-heavy} phase has relatively long input sequences and short output sequences; a \textit{decode-heavy} phase has relatively short input sequences and long output sequences; a \textit{balanced-short} phase has short inputs and short outputs; and a \textit{stable-mixed} or \textit{stable-decode-heavy} phase denotes a later stable period whose token-shape profile remains similar across consecutive timestamps.

\noindent\textbf{DistServe.}
The DistServe evaluation uses two traces sampled from \textit{ShareGPT} and \textit{LongBench}, respectively. We divide each trace into six workload phases, where each timestamp corresponds to the beginning of a workload phase. Each of the six \textit{ShareGPT} phases contains $5120$ requests, and each \textit{LongBench} phase contains $1728$ requests. Specifically, the \textit{ShareGPT} trace contains one prefill-heavy phase, one decode-heavy phase, one short balanced phase, and then three stable mixed phases. The \textit{LongBench} trace contains two prefill-heavy phases, one decode-heavy transition phase, and then three stable decode-heavy phases. The phase-wise mean token-shape profiles are listed in Table~\ref{tab:distserve-hexgen-traces}.

\noindent\textbf{HexGen.}
The HexGen evaluation uses the same phase-wise token-shape profiles as DistServe. The difference is workload intensity: each workload phase in \textit{ShareGPT} trace contains 9216 requests, and each workload phase in \textit{LongBench} contains 3110 requests. Thus, the prefill and decode lengths in Table~\ref{tab:distserve-hexgen-traces} remain unchanged, while the effective workload size is scaled up.

\begin{table}[t]
\centering
\caption{%
\textit{Phase-wise workload profiles used by DistServe and HexGen:} The mean prefill length and decode length in each phase are presented.
}
\label{tab:distserve-hexgen-traces}
\resizebox{0.44\linewidth}{!}{%
\begin{tabular}{c|c|c|r|r}
\hline
\textbf{Trace} & \textbf{TS} & \textbf{Phase} & \textbf{Pref} & \textbf{Dec} \\
\hline
\multirow{6}{*}{ShareGPT}
& ts0 & Prefill-heavy       & 1231.9 & 14.0  \\
& ts1 & Decode-heavy        & 535.2  & 545.4 \\
& ts2 & Balanced-short      & 549.0  & 17.5  \\
& ts3 & Stable-mixed        & 1093.9 & 290.3 \\
& ts4 & Stable-mixed        & 1100.6 & 292.4 \\
& ts5 & Stable-mixed        & 1096.8 & 288.5 \\
\hline
\multirow{6}{*}{LongBench}
& ts0 & Prefill-heavy       & 2035.2 & 4.8   \\
& ts1 & Prefill-heavy       & 2036.8 & 3.2   \\
& ts2 & Decode-heavy        & 1596.6 & 372.6 \\
& ts3 & Stable-decode-heavy & 1604.8 & 373.1 \\
& ts4 & Stable-decode-heavy & 1553.7 & 396.7 \\
& ts5 & Stable-decode-heavy & 1582.4 & 386.8 \\
\hline
\end{tabular}%
}
\end{table}

\noindent\textbf{SpotServe.}
This evaluation uses two MAF-derived traces, each combining hidden workload-pattern changes with a time-varying elastic cluster-size schedule. Both traces are extracted from a real Azure Functions application window and then mapped into an elastic cluster schedule ranging from $24$ to $64$ GPUs. In the current construction, the two traces share the same planner-visible cluster schedule. Table~\ref{tab:spotserve-workloads} lists the hidden workload phases, and Table~\ref{tab:spotserve-cluster} lists the shared cluster-size schedule.

\begin{table}[t]
\centering
\caption{%
\textit{SpotServe traces}: hidden workload phases used in the SpotServe evaluation.
}
\label{tab:spotserve-workloads}
\resizebox{0.41\linewidth}{!}{%
\begin{tabular}{c|c|c|r|r}
\hline
\textbf{Trace} & \textbf{TS} & \textbf{Phase} & \textbf{Pref} & \textbf{Dec} \\
\hline
\multirow{4}{*}{MAF Trace 1}
& ts0 & Decode-heavy  & 512  & 1024 \\
& ts1 & Mixed         & 2048 & 256  \\
& ts2 & Prefill-heavy & 4096 & 128  \\
& ts3 & Mixed-stable  & 2048 & 256  \\
\hline
\multirow{4}{*}{MAF Trace 2}
& ts0 & Prefill-heavy & 4096 & 128  \\
& ts1 & Mixed         & 2048 & 256  \\
& ts2 & Decode-heavy  & 512  & 1024 \\
& ts3 & Mixed-stable  & 2048 & 256  \\
\hline
\end{tabular}%
}
\end{table}

\begin{table*}[t!]
\centering
\caption{%
\textit{Shared SpotServe cluster schedule}: planner-visible elastic cluster sizes for both MAF traces. Each control bucket spans $3$ seconds.
}
\label{tab:spotserve-cluster}
\resizebox{0.95\linewidth}{!}{%
\begin{tabular}{c|cccccccccccc}
\hline
\textbf{Time Interval (s)} 
& [0,3) & [3,6) & [6,9) & [9,12) & [12,15) & [15,18) & [18,21) & [21,24) & [24,27) & [27,30) & [30,33) & [33,36) \\
\hline
\textbf{Cluster Size}
& 24 & 25 & 26 & 27 & 29 & 30 & 32 & 33 & 36 & 38 & 42 & 45 \\
\hline
\textbf{Time Interval (s)} 
& [36,39) & [39,42) & [42,45) & [45,48) & [48,54) & [54,57) & [57,60) & [60,63) & [63,66) & [66,69) & [69,75) & [75,78) \\
\hline
\textbf{Cluster Size}
& 48 & 51 & 54 & 55 & 60 & 63 & 62 & 64 & 61 & 62 & 60 & 57 \\
\hline
\textbf{Time Interval (s)} 
& [78,81) & [81,87) & [87,90) & [90,93) & [93,96) & [96,99) & [99,105) & [105,108) & [108,114) & [114,117) & [117,120) & \\
\hline
\textbf{Cluster Size}
& 56 & 54 & 55 & 53 & 51 & 50 & 49 & 47 & 45 & 44 & 43 & \\
\hline
\end{tabular}%
}
\end{table*}

\section{Discussion and Limitation}
\label{appendix:discussion}

\noindent\textbf{LLM API stability and cost.} The evolutionary process relies on LLM API calls for candidate mutation, making it sensitive to API availability and latency. Unstable API endpoints can stall individual evolution cycles, prolonging the time before an improved policy is deployed. Additionally, LLM inference introduces non-trivial monetary cost---commercial APIs such as Claude charge considerable per-token fees that may be prohibitive for continuous evolution in production. To mitigate this, \sys is designed to be model-agnostic: Our current deployment uses open-source models, making operational cost substantially more manageable while maintaining competitive synthesis quality. This also demonstrates that our approach does not rely on frontier-class closed-source LLMs to be effective.

\vspace{0.25em}
\noindent\textbf{LLM capability and hallucination.} The quality of evolved policies is inherently bounded by the code-generation capability of the underlying LLM. Stronger models generally produce more effective mutations and converge faster, while weaker models may require more iterations. Moreover, LLM hallucinations can generate syntactically valid but semantically incorrect policies. Our multi-level timeout hierarchy (\S\ref{sec:impl}) and automated evaluation (\S\ref{subsec:evaluator}) serve as safeguards by discarding candidates that crash, diverge, or degrade fitness, but hallucination rates at production scale remain a factor that influences evolution efficiency.

\vspace{0.25em}
\noindent\textbf{Snapshot window configuration.} The length of the runtime trace snapshot presents a trade-off: An overly long window aggregates information across multiple workload regimes, diluting the signal of the current operating state and potentially producing policies optimized for an averaged rather than present condition. Conversely, an overly short window may be dominated by transient fluctuations, causing the evolved policy to overfit to momentary noise rather than sustained trends. In our current implementation, the window size is a configurable parameter; adaptive window sizing based on detected workload volatility is a promising direction for future work.

\vspace{0.25em}
\noindent\textbf{Evolution time overhead.} Each evolution cycle incurs wall-clock time for candidate generation, evaluation, and selection. We address this through two mechanisms: (i) Warm-start re-evolution (\S\ref{sec:impl}), which reduces convergence time by up to 83\% when consecutive snapshots reflect gradual shifts, and (ii) parallel candidate evaluation, which accelerates the evaluation stage by up to 8$\times$ with 10$\times$ more concurrent CPU threads. Further reduction through speculative evaluation or early termination of unpromising candidates remains an avenue for future optimization.

\vspace{0.25em}
\noindent\textbf{Adaptability to diverse serving objectives.} Our execution model and evaluator are formulated around end-to-end trace completion time, but this choice is not fundamental to the architecture. The policy interface (\texttt{should\_reschedule} and \texttt{schedule}) and the three-phase interval structure remain unchanged regardless of the objective. To optimize for alternative metrics such as P95 latency or SLO attainment, only the evaluator's aggregation logic needs modification: Instead of summing per-interval costs into $\text{T}_{\text{total}}$, the simulator tracks per-request latency across the trace replay and the evaluator computes the target metric (e.g., the 95th-percentile latency or the fraction of requests meeting a latency threshold) as the scalar fitness score. Since the synthesis workflow treats the fitness score as a black-box scalar, this substitution requires no changes to the mutation, selection, or deployment mechanisms, enabling \sys to optimize for the serving metric most relevant to a given deployment scenario.

\section{Extended Related Work}
\label{appendix:relatedwork}

\textbf{LLM serving systems.} Recent works have proposed several systems and methods for highly efficient LLM serving~\cite{zhang2025efficient,wang2025thinking,contributors2023lmdeploy,jiang2025cascadia,peng2025hexgen,jiang2025hexgen,jiang2025thunderserve,jiang2026boute,jiang2026oserve,tong2025parallax,he2026efficient}. AlpaServe~\cite{li2023alpaserve} optimizes system service level objectives (SLOs) by utilizing data and model parallelism; DistServe~\cite{zhong2024distserve} and Splitwise~\cite{patel2024splitwise} split the prefill and decoding phases onto separate GPUs to eliminate interference between them; SarathiServe~\cite{agrawal2024taming} optimizes request batching through prefill chunking to mitigate interference between the prefill and decoding stages; vLLM~\cite{kwon2023efficient} introduces PagedAttention for efficient memory management, enabling higher batch sizes and improved throughput; FastServe~\cite{wu2023fast} employs a preemptive scheduling mechanism that prioritizes shorter jobs to minimize job completion time. These works primarily focus on improving serving performance through techniques such as batching optimization, phase separation, and memory management. In contrast, our work focuses on deployment scheduling with heterogeneous LLMs and GPU types to achieve cost-efficient LLM serving.

\textbf{Parallelism strategies.} Training and serving LLMs that exceed the memory capacity of a single GPU requires a combination of parallelism strategies. Data parallelism replicates the full model across devices and partitions training batches or distributes inference requests among replicas, scaling throughput but requiring each device to hold the entire model~\cite{li2023alpaserve}. Tensor parallelism partitions individual operations — such as the large matrix multiplications in attention and feed-forward layers — across devices within a node, reducing per-GPU memory usage and per-step or per-request latency at the cost of frequent all-reduce communication~\cite{shoeybi2019megatron,miao2022galvatron}. Pipeline parallelism divides the model into sequential stages across devices, enabling larger models to be trained or served by overlapping the execution of different micro-batches across stages, though it introduces pipeline bubbles and increased latency~\cite{huang2019gpipe}. Hybrid parallelism combines all three approaches to efficiently scale massive models while jointly balancing latency, throughput, and memory constraints~\cite{jiang2022osdp,guoliangefficient,wang2024improving,yan2024hexiscale}.

\end{document}